\documentclass[fleqn,usenatbib,nofootinbib]{mnras}
\usepackage{footnote}
\usepackage{newtxtext,newtxmath}
\usepackage[T1]{fontenc}
\usepackage{threeparttable}

\DeclareRobustCommand{\VAN}[3]{#2}
\let\VANthebibliography\thebibliography
\def\thebibliography{\DeclareRobustCommand{\VAN}[3]{##3}\VANthebibliography}

\usepackage{graphicx}	
\usepackage{amsmath}	
\usepackage{multirow}

\usepackage{bm} % For bold math (esp of lower greek letters)
\usepackage{dcolumn}% Align table columns on decimal point
\usepackage{color}
\usepackage{multicol}
\usepackage{ulem}

\usepackage{comment}
\usepackage{mathrsfs}
\usepackage{amsfonts}
\usepackage{varioref}
\usepackage{csquotes}

\usepackage{float}
\usepackage[caption=false]{subfig}

\def \Msun {M_\odot}

\def \gc {globular cluster}

\def \ET {Einstein telescope}
\def \CE {Cosmic explorer}

\labelformat{chapter}{Chapter (#1)} 
\labelformat{section}{Section (#1)} 
\labelformat{subsection}{Section (#1)} 
\labelformat{subsubsection}{Section (#1)}
\labelformat{subsubsubsection}{Section (#1)}
\labelformat{equation}{Eq.~(#1)} 
\labelformat{figure}{Fig.~(#1)} 
\labelformat{subfigure}{Fig.~(\thefigure#1)} 
\labelformat{table}{Tab.~(#1)} 
\labelformat{appendix}{Appendix #1}%%%%%%%%%%%%%%%%%Macros for some shortforms %%%%%%%%%%%%%%%%%%%%%%%%%%%%%%%%%%%%%%%%%%%%%%%%%
\title[Detectability of GW from hyperbolic encounters of BHs]{Gravitational Wave observatories may be able to detect hyperbolic encounters of Black Holes}

\author[S. Mukherjee et al.]{
Sajal Mukherjee,$^{1}$\thanks{E-mail: sajal@iucaa.in}
Sanjit Mitra,$^{1}$
and Sourav Chatterjee $^{2}$
\\
$^{1}$ Inter-University Centre for Astronomy and Astrophysics (IUCAA), Post Bag 4, Pune 411007, India\\
$^{2}$ Tata Institute of Fundamental Research, Homi Bhabha Road, Navy Nagar, Colaba, Mumbai 400005, India\\ 
}
\date{\today}

\pubyear{2021}

\begin{document}

\label{firstpage}
\pagerange{\pageref{firstpage}--\pageref{lastpage}}
\maketitle
%%%%%%%%%%%%%%%%%%%%%%%
\begin{abstract}
%%%%%%%%%%%%%%%%%%%%%%%%
Gravitational Wave (GW) astronomy promises to observe different kinds of astrophysical sources. Here we explore the possibility of detection of GWs from hyperbolic interactions of compact stars with ground-based interferometric detectors. It is believed that a bound compact cluster, such as a globular cluster, can be a primary environment for these interactions. We estimate the detection rates for such events by considering local geometry within the cluster, accounting for scattering probability of compact stars at finite distances, and assuming realistic cluster properties guided by available numerical models, their formation times, and evolution of stars inside them. We find that, even in the conservative limit, it may be possible to detect such black hole encounters in the next few years by the present network of observatories with the ongoing sensitivity upgrades and one to few events per year with the next generation observatories. In practice, actual detection rates can significantly surpass the estimated average rates, since the chances of finding outliers in a very large population can be high. Such observations (or, no observation) may provide crucial constraints to estimate the number of isolated compact stars in the universe. These detections will be exciting discoveries on their own and will be complimentary to observations of binary mergers bringing us one step closer to address a fundamental question, how many black holes are there in the observable universe. 
%%%%%%%%%%%%%%%%%%%%%%%
\end{abstract}
%%%%%%%%%%%%%%%%%%%%%%%%

%%%%%%%%%%%%%%%%%%%%%%%%%%%%%%%%%%%%%%%%%%%%%%%%%%

%%%%%%%%%%%%%%%%% BODY OF PAPER %%%%%%%%%%%%%%%%%%
\section{Introduction}
%%%%%%%%%%%%%%%%%%%%%%%%%%%%%%%%%%%%%%%%%%%%%%%%%%%
With the advancement in detecting two body interactions like binary mergers, there is a growing interest to study scattering events pertaining to GW astronomy (\citealt{Cho:2018upo}). Not only these studies are useful to unveil yet unknown features of populations of astrophysical sources, but also hint interesting bulk properties of the galaxies and clusters (\citealt{OLeary:2008myb}). The remarkable success story of GW astronomy (\citealt{Abbott:2016blz,TheLIGOScientific:2017qsa,Abbott:2017gyy,LIGOScientific:2018mvr,Abbott:2020uma,Abbott:2020khf}) has boosted these searches, and some of the recent works can be found in (\citealt{Huerta:2017kez,Gondan:2017wzd,Vines:2018gqi,jakobsen2021gravitational,saketh2021conservative,Jakobsen:2021smu}). Even though the current analyses and waveform modeling are incapable of detecting such events, an enormous amount of progress has been achieved in recent years to understand eccentric and highly eccentric interactions between stars or black holes (\citealt{DeVittori:2012da,Grobner:2020fnb}). These events are studied in connection to laser interferometric detectors (\citealt{Tiwari:2015gal,Zevin:2018kzq,Capozziello:2008mn}), and are relevant for the current and upcoming GW detectors such as Advanced LIGO (\citealt{TheLIGOScientific:2014jea,ccite}), Virgo (\citealt{TheVirgo:2014hva,VIRGO_cite}), KAGRA (\citealt{Aso:2013eba,KAGRA_cite}), LIGO-India (\citealt{LIGOM1100296-v2,LIGO_India_cite}), Einstein Telescope (ET) (\citealt{Punturo:2010zza,ET:cite}), and Cosmic Explorer (CE) (\citealt{Evans:2016mbw,CE:cite}).

The possible detection of the hyperbolic encounters by the ground based GW detectors set the backstage of motivation of our article. With these events likely to be detected in near future, it is of significant interest to study major components of scattering events. In particular, any such detection would crucially depend on the orbital model and the background environment where these encounters usually take place. It is suspected and typically argued that clusters of stars with high population density can be most favorable to host these interactions (\citealt{dymnikova1982bursts,Kocsis:2006hq}). Some of the other important channels where hyperbolic interactions are likely to be present include three body problems (\citealt{Trani:2019nij}), active galactic nuclei (\citealt{OLeary:2008myb}) or primordial BHs (\citealt{Garcia-Bellido:2017knh}). However, in the present article, we will only focus on the hyperbolic interactions taking place inside a star cluster and study various components of the hyperbolic scattering problem. Typically speaking, clusters are loosely divided into two categories --- open and closed; and here, we will only consider the closed or otherwise known as the \gc~(GC). These clusters contain a large population of stars within a radius of a few parsecs, and consist of a dense core where the majority of these close counters occur (\citealt{heggie_hut_2003}). It turns out to be notoriously difficult to model these interactions accurately (\citealt{2009ApJ...690.1370M,10.1111/j.1365-2966.2009.15880.x,PhysRevD.93.084029}). In the present study, we employ a non-relativistic model to study hyperbolic encounters occurring inside these clusters for simplicity and gaining insight.

In (\citealt{dymnikova1982bursts,Kocsis:2006hq,Heggie:1995yk}), scattering/hyperbolic encounters inside a \gc~are discussed in connection with the GW detectors. However, these interactions are described as the classical scattering problem that we are familiar with, where the object is assumed to be coming from the spatial infinity. This framework requires two independent parameters to model the interaction, namely, velocity at infinity, and impact parameter. Nonetheless, from realistic perspective, while describing scattering incidents inside a close cluster with $R_{\rm c} (\sim \mathcal{O}(1) $~pc), where $R_{\rm c}$ is the radius of the cluster, the initial distance, $r_i$, is always bounded as $r_i \leq R_{\rm c}$. To comply with the conservation of energy and momentum, this change may affect the overall dynamics of the two body scattering problem. Within the Newtonian domain, we aim to study how the hyperbolic encounter changes in case we impose such restriction on $r_i$. In particular, we investigate the effects of relaxing the assumption of projectiles from infinity in the traditional scattering calculation, by taking into account local effects inside the cluster. This would involve changing the initial conditions of the interaction and introducing different sets of parameters accountable for that. In our case, we found that the initial distance and angle can be useful to study the scattering problem and obtain various orbital entities, like eccentricity, GW radiation and event rates. This, in fact, is one of the primary objectives of the present study.

The other part of the paper motivates searches of scattering events from the third generation detector's perspective, such as \ET~and \CE, and also upgraded versions of the advanced LIGO, such as LIGO A+ and LIGO-Voyager. These detectors typically consist of sensitivity larger than advanced LIGO, and can be probed to detect GW signal from high redshited sources. In (\citealt{sathyaprakash2019multimessenger,kalogera2019deeper,Evans:2016mbw}), the comparison between various detectors and their sensitivities is given in depth. However, to the best of our knowledge, the rates of hyperbolic interaction as observed by these detectors have not yet been explored in literature,  which is another key objective of the present study. Here we provide a detailed discussion on how event rates vary for different detectors with different sensitivity.

The rest of the manuscript is organized as follows: In \ref{sec:Backbone}, we introduce the backbone calculations of the present model while incorporating the local effects inside the cluster. The main results are given in \ref{sec:Event_Rate}, where we estimate the event rate for these interactions. Finally, we conclude the article with a brief remark in \ref{sec:Discussion}.
%%%%%%%%%%%%%%%%%%%%%%%%%%%%%%%%%%%%%%%%%%%%%%%%%%%%%
\section{The orbital model and backbone calculations}\label{sec:Backbone}
\begin{figure}
\centering
\includegraphics[width=0.45\textwidth]{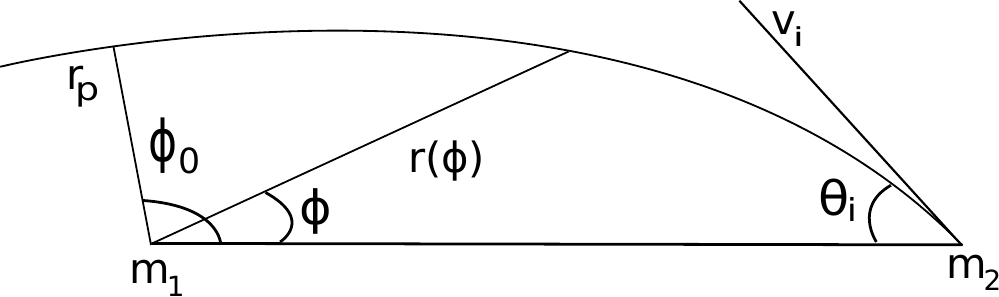}
\caption{Schematic diagram we use here to study the scattering incidents of masses $m_1$ and $m_2$ with initial relative speed $v_i$ at an angle $\theta_i$ with respect to the separation vector. Depending on  $\theta_i$, the power radiation is dictated. The trajectory is described by the radial distance $r$ as a function of the azimuthal angle $\phi$. $\phi_0$ and $r_p$ denote the angle at periapsis and periapsis distance respectively.}
\label{fig:trajectory}
\end{figure}

%%%%%%%%%%%%%%%%%%%%%%%%%%%%%%%%%%%%%%%%%%%%%%%%%%%%%
In this section, we will briefly go through the scattering model that we choose to work with. Unlike the usual scattering problem, where the objects are infinity apart initially, here we assume that the initial distance, $r_i$, is taken to be within the radius of the cluster. Besides, two other parameters, namely, initial angle between binary components, $\theta_i$, and initial velocity, $v_i$, are essential to model the hyperbolic interaction. With a precise knowledge of these parameters, it is possible to define the orbits and obtain quadrupole moment of the binary, which leads to other entities like energy radiation and GW waveform. Out of these parameters, $r_i$ always follows $r_i \leq R_{\rm c}$ and $v_i$ can be estimated with sufficient accuracy by using the \textit{virial theorem}, but the angular parameter $\theta_i$ can be arbitrary, and needs to be constrained precisely to measure the event rate. The maximum $\theta_i$ is obtained by assuming that the signal to noise ratio has to exceed threshold value; while the minimum is evaluated by assuming that the \textit{periastron distance} is always greater than twice of the \textit{schwarzschild radius} ($r_{\rm s}$) of an individual binary component. In \ref{fig:trajectory}, we illustrate the basic model, which is a classic two-body problem, and we use the non-relativistic solution by invoking conservation of energy and angular momentum. 
%%%%%%%%%%%%%%%%%%%%%
%%%%%%%%%%%%%%%%%% 
We start with the conservation of total energy ($E$) and momentum ($L$) per unit mass, which indicates  that the energy radiated in GW is negligible to make any correction on the orbital dynamics:
%%%%%%%%%%%%%%%%%%%%%%%%%%%%%%
\begin{equation}
E=E_{k}+\Phi(r)=\dfrac{1}{2}\left\{\Bigl(\dfrac{dr}{dt} \Bigr)^2+r^2 \Bigl(\dfrac{d\phi}{dt} \Bigr)^2\right\}-\dfrac{G M}{r},
\end{equation}
%%%%%%%%%%%%%%%%%%%%%%%%%%%%%%
where, $r$ and $\phi$ are given as the radial and angular coordinates respectively, $E_{k}$ is the kinetic energy, $\Phi(r)$ is the potential energy, and $M$ is the total mass given by $M=m_1+m_2$. With the substitution, $r=1/u$, we can rewrite the above equation as follows:
%%%%%%%%%%%%%%%%%%%%%%%%%%%%%
\begin{equation}
\dfrac{d^2u}{d\phi^2}+u = -\dfrac{G M}{L^2},
\end{equation}
%%%%%%%%%%%%%%%%%%%
which is the motion of a particle in a central force field, and the conserved momentum is given by, $L=r_i v_i \sin\theta_i$. The above has a solution of the following form:
%%%%%%%%%%%%%%%%%
\begin{equation}
u(\phi)=\dfrac{1}{r(\phi)}=C \cos(\phi-\phi_0)+\dfrac{GM}{L^2},
\label{eq:radial_distance}
\end{equation}
%%%%%%%%%%%%%%%%%
with $\phi_0$ is the angle at the periastron. In the above, $C$ being the integration constant, and needs to be evaluated from the initial conditions. With the initial condition, $\phi=0$, $r=r_i$, we have
%%%%%%%%%%%%%%%%%%%
\begin{equation}
\dfrac{1}{r_i}=C \cos\phi_0+\dfrac{GM}{L^2},
\end{equation}
%%%%%%%%%%%%%%%%%%%
and $\phi=\phi_0$, $r=r_p=r_{\rm min}$
%%%%%%%%%%%%%%%%%%%
\begin{equation}
C=\dfrac{1}{r_p}-\dfrac{GM}{L^2}.
\label{eq:ri_rp}
\end{equation}
%%%%%%%%%%%%%%%%%%%
From the above two equations, $r_p$ can be written in terms of the initial conditions as follows:
%%%%%%%%%%%%%%%%%%
\begin{equation}
r_p=\dfrac{r^2_i v^2_i \cos\phi_0}{r_i v^2_i+GM (\cos\phi_0-1)\csc^2\theta_i}.
\label{eq:r_p}
\end{equation}
%%%%%%%%%%%%%%%%%%%
The above equation gives a direct relation between the initial conditions and $r_p$. Finally, the radial distance given in \ref{eq:radial_distance} becomes
%%%%%%%%%%%%%%%%%%%
%%%%%%%%%%%%%%%%%%%
\begin{equation}
r(\phi)=\dfrac{{L^2}/{GM}}{1+\left({L^2}/{GMr_p}-1 \right)\cos(\phi-\phi_0)}.
\label{eq:r_rp}
\end{equation}
%%%%%%%%%%%%%%%%%%%
The above is the final equation for the radial distance as a function of $\phi$. As we may infer, the eccentricity $e$, and semi-latus rectum $p$, can now be given as
%%%%%%%%%%%%%%%%
\begin{equation}
p=\dfrac{GM}{L^2}, \quad \text{and} \quad e=\dfrac{L^2}{GMr_p}-1.
\end{equation}
%%%%%%%%%%%%%%%% 

Let us now use the following relation which relates initial velocity $v_i$ with $r_i$:
%%%%%%%%%%%%%%%%%
\begin{eqnarray}
v_i^2 &=& \left(\dfrac{dr}{dt}\right)^2_{r=r_i}+r_i^2 \left(\dfrac{d\phi}{dt}\right)^2_{r=r_i}=\left(\dfrac{dr}{dt}\right)^2_{r=r_i}+v_i^2 \sin^2\theta_i, \nonumber \\
\end{eqnarray}
%%%%%%%%%%%%%%%%%
and, we have 
%%%%%%%%%%%%%%%%%
\begin{equation}
v_i^2 \cos^2\theta_i=\left(\dfrac{dr}{dt}\right)^2_{r=r_i}.
\label{eq:vi_r_i}
\end{equation}
%%%%%%%%%%%%%%%%%
Introducing the following relation, 
%%%%%%%%%%%%%%%%
\begin{eqnarray}
\left(\dfrac{dr}{dt} \right)_{r=r_i}=\left(\dfrac{dr}{du}\dfrac{du}{d\phi}\dfrac{d\phi}{dt}\right)_{r=r_i} = r_{i}v_{i}\sin\theta_i C \sin\phi_0, \nonumber \\
\end{eqnarray}
%%%%%%%%%%%%%%%%
and substituting it back to \ref{eq:vi_r_i}, we gather
%%%%%%%%%%%%%%%%
\begin{equation}
v^2_i \cos^2\theta_i= r^2_i v^2_i \sin^2\theta_i C^2 \sin^2\phi_0,
\end{equation}
%%%%%%%%%%%%%%%%
which, upon using \ref{eq:ri_rp}, becomes
%%%%%%%%%%%%%%%%
\begin{equation}
\sin\phi_0=\dfrac{\cot\theta_i}{r_i}\left\{\dfrac{1}{r_p}-\dfrac{GM}{L^2}\right\}^{-1}.
\end{equation}
%%%%%%%%%%%%%%%%
By employing \ref{eq:r_p}, we can rewrite the above as follows:
%%%%%%%%%%%%%%%
\begin{equation}
\tan\phi_0=\dfrac{r_i v_i^2 \cot\theta_i}{r_i v^2_i-GM \csc^2\theta_i}.
\end{equation}
%%%%%%%%%%%%%%%
%%%%%%%%%%%%%%%%%%%%%%%%%%%%%%%%%%%%%%%%%%%%%%%%
The above gives a relation between the initial conditions and angle at periastron. Finally, various orbital quantities such as eccentricity and $\phi_0$ can be written in a more compact form as follows:
%%%%%%%%%%%%%%%%%%
\begin{eqnarray}
& & e^2 = 1+\dfrac{L^2 vi^2}{G^2M^2}\left\{1-\left(\dfrac{2GM}{v_i^2}\right) \dfrac{1}{r_i}\right\}, \nonumber \\
& &r_p = \dfrac{L^2 \cos\phi_0}{L^2/r_i-GM(1-\cos\phi_0)}=\dfrac{L^2}{GM(1+e)},\nonumber \\
& & \tan\phi_0  = \dfrac{Lv_i \cos\theta_i}{L^2/r_i -GM} = -\dfrac{L v_i}{GM}\left\{\dfrac{{\cos \theta_i}}{1-L^2/(GM r_i)}\right\}. \nonumber \\
\label{eq:scattering_our}
\end{eqnarray}
%%%%%%%%%%%%%%%%%%
Moreover, we choose to work within the domain $0<\theta_i < \pi/2$ and $\pi \leq \phi_0 < 0$. For $\theta=\pi/2$, \ref{eq:scattering_our} will give $\phi_0=0$, and substitute it back to \ref{eq:r_p}, we have $r_p=r_i$. This indicates that there is no interaction, and therefore, we exclude this possibility on physical ground.
%%%%%%%%%%%%%%%%%%%%%%%%%%%%%%%%%%%%%%%%%%%%%%%%%%%%%
\begin{figure}
\centering
\includegraphics[width=0.45\textwidth]{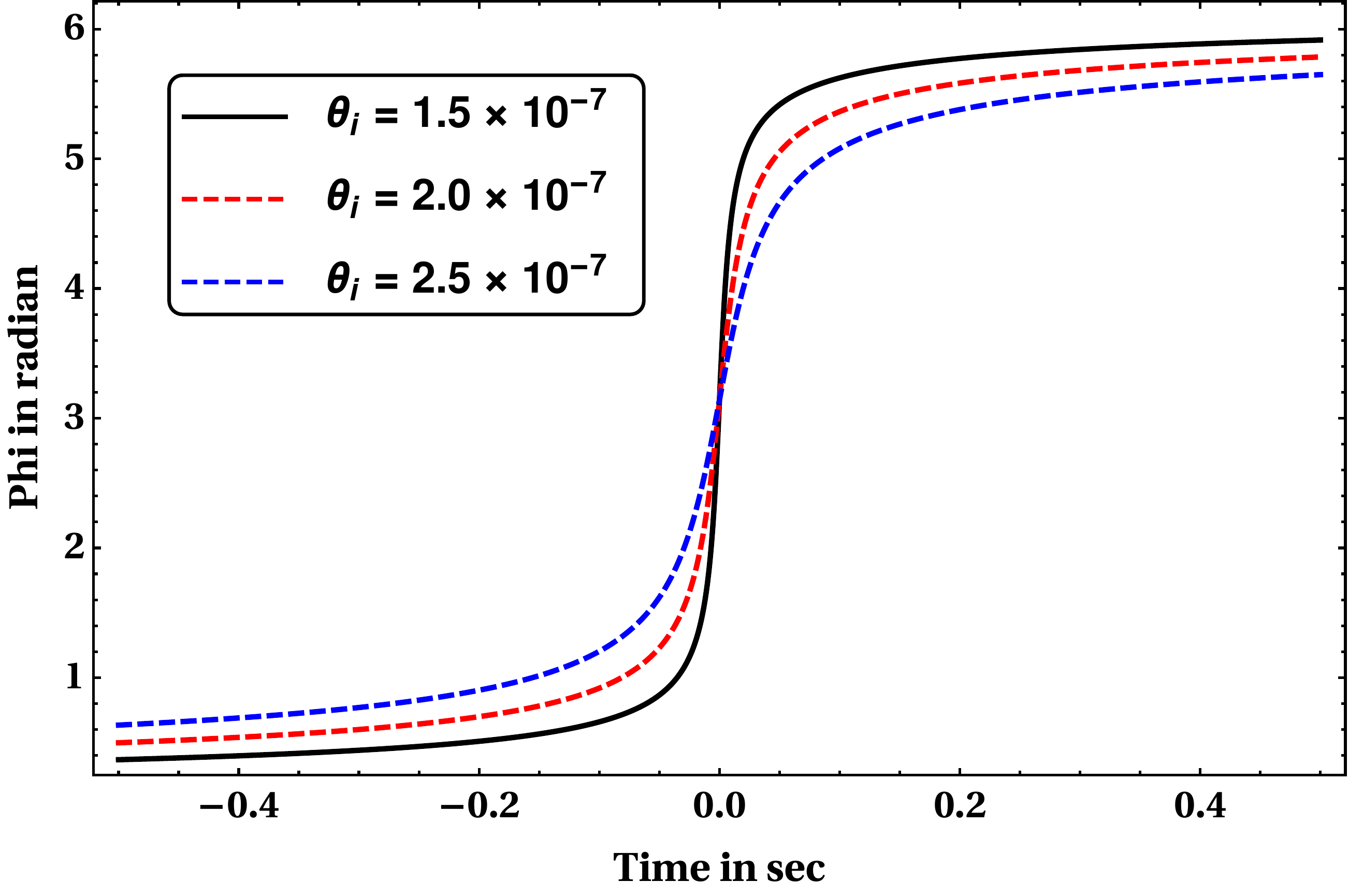}
\includegraphics[width=0.45\textwidth]{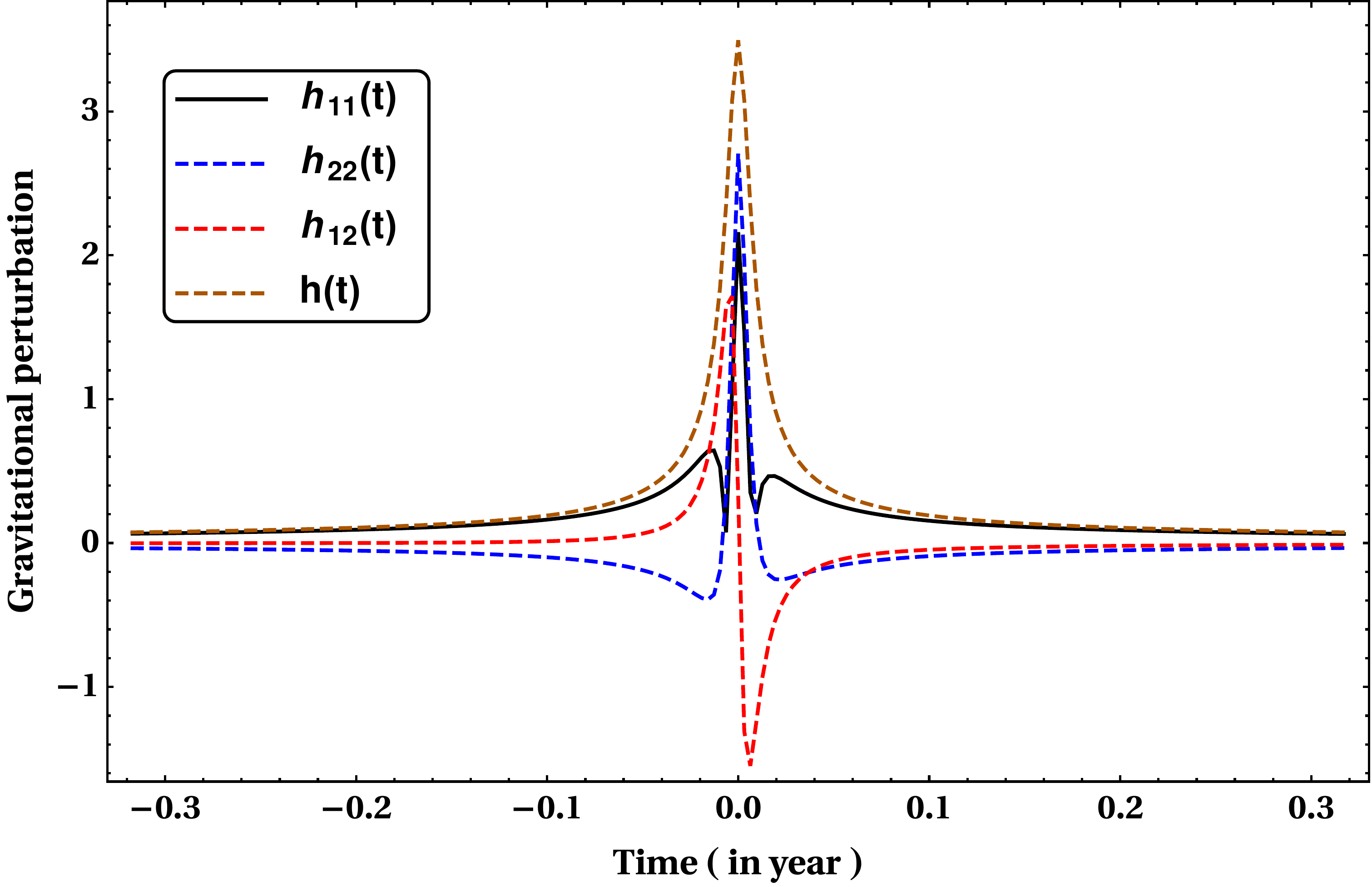}
\caption{In the above figure, we demonstrate the change in angular coordinate as a function of time, while we set $r_i=1$pc. For smaller value of $\theta_i$, the interaction becomes stronger and takes lesser time compare to larger values of $\theta_i$. The  mass of an individual binary component is $m_1=m_2=10 \Msun$. Corresponding perturbations in the space-time metric (scaled with $10^{24}$) are shown for $r_i=1 \text{pc}$, $m_1=m_2=10 M_{\odot}$, $\theta_i=10^{-4}$ and the eccentricity is $1.01378$. The black curve is given as $h_{11}(t)$, while the blue and red corresponds to $h_{22}(t)$ and $h_{12}(t)$ respectively. Finally, the brown curve gives the value of $h(t) = \left\{h_{11}(t)^2+h_{22}(t)^2+2 h_{12}(t)^2 \right\}^{1/2}$.}
\label{fig:collage}
\end{figure}
%%%%%%%%%%%%%%%%%%
Depending on the initial conditions, the trajectory changes with time, and shown in \ref{fig:collage}. As mentioned earlier, the $v_i$ is obtained by employing the virial theorem (\citealt{longair2007galaxy})
%%%%%%%%%%%%%%%%%
\begin{equation}
v_i=\sqrt{\dfrac{G \overline{M}}{3 \overline{R}}}=\sqrt{\dfrac{G n_{\rm star}m}{3 R_{\rm c}}},
\label{eq:virial_vel}
\end{equation}
%%%%%%%%%%%%%%%%%
where, we have used the average mass of the cluster to be total number of stars, $n_{\rm star}$, multiplied with average mass of a star ($m$), i.e., $\overline{M}=n_{\rm star}m$, and average distance to be radius of the cluster $R_{\rm c}$.

As the first check, we explore a limit where the present model would merge with the existing literature. With $r_i \rightarrow \infty$, and $\theta_i \rightarrow 0$, such that the angular momentum $L=r_i v_i \sin\theta_i$ remains finite, we gather
%%%%%%%%%%%%%%%%%%
\begin{eqnarray}
& & e^2 = \sec^2\phi_0=1+\dfrac{L^2 v_i^2}{G^2M^2}, \nonumber \\
& &r_p = \dfrac{-L^2 \cos\phi_0}{GM(1-\cos\phi_0)}=\dfrac{L^2}{GM(1+e)},\nonumber \\
& & \tan\phi_0  = -\dfrac{L v_i}{GM},
\label{eq:scattering_usual}
\end{eqnarray}
%%%%%%%%%%%%%%%%%%
which represents the usual orbital parameters to describe the classical hyperbolic interaction (\citealt{DeVittori:2012da}).
%%%%%%%%%%%%%%%%%%%%%%%%%%%%%%%%%%%%%%%%%%%%%%%
\section{Estimation of the event rate}\label{sec:Event_Rate}
%%%%%%%%%%%%%%%%%%%%%%%%%%%%%%%%%%%%%%%%%%%%%%%%%%%%
%%%%%%%%%%%%%%%%%%%%%
In this section, we will evaluate the rate for scattering/hyperbolic encounters as a function of initial conditions, namely $r_i$ and $\theta_i$. The formalism is loosely based on the scattering of gas molecules inside a closed container. For a fixed initial distance, $r_i$, we can estimate the solid angle $\Omega$ (as a function of $\theta_{\rm max}(r_i)$ and $\theta_{\rm min}(r_i)$) within which all scattering events produce detectable signals:
%%%%%%%%%%%%%%%%%%%%%
\begin{equation}
\Omega=2\pi\int^{\theta_{\rm max}(r_i)}_{\theta_{\rm min}(r_i)}\sin\theta d\theta=\pi \left[\theta_{\rm max}(r_i)^2-\theta_{\rm min}(r_i)^2\right].
\end{equation}
%%%%%%%%%%%%%%%%%%%%%
We assume that the objects are distributed isotropically inside the cluster. The probability of selecting this fraction of particles falls within the above solid angle is 
%%%%%%%%%%%%%%%%%%%%%
\begin{equation}
P_{\rm theta}=\dfrac{1}{4} \left[\theta_{\rm max}(r_i)^2-\theta_{\rm min}(r_i)^2\right].
\end{equation}
%%%%%%%%%%%%%%%%%%%%%
For an individual object inside the cluster, the event rate or number of events per unit time becomes:
%%%%%%%%%%%%%%%%%%%%%
\begin{equation}
P_{\rm indv}=\int^{R_{\rm c}}_{R_{\rm min}}\Bigl(\dfrac{P_{\rm theta}}{t_{\rm col}}\Bigr) 4 \pi r_i^2 n_{\rm s} dr_i,
\end{equation}
%%%%%%%%%%%%%%%%%%%%% 
where, $t_{\rm col}$ is the average time of a collision, and can be written as $t_{\rm col} \le r_i/v_i$, $n_{\rm s}$ is the uniform volume density of the stars, i.e., $n_{\rm s}=3n_{\rm star}/(4 \pi R_{\rm c}^3)$. The above expression can be simplified as
%%%%%%%%%%%%%%%%%%%%
\begin{equation}
P_{\rm indv}=\dfrac{3v_i n_{\rm star}}{4 R_{\rm c}^3}\int^{R_{\rm c}}_{R_{\rm min}}\bigl[\theta_{\rm max}(r_i)^2-\theta_{\rm min}(r_i)^2\bigr]r_i dr_i.
\end{equation}
%%%%%%%%%%%%%%%%%%%%
Remarkably it turns out that both the expressions $b_{\rm max}=r_{i}\theta_{\rm max}(r_i)$ and $b_{\rm min}=r_{i}\theta_{\rm min}(r_i)$, are independent of the initial distance ($r_i$), and remains conserved for a variety of initial distances. It seems that the conservation of total angular momentum engineers to produce this incident.
%%%%%%%%%%%%%%%%%%%%%%%%%%%%%
%%%%%%%%%%%%%%%%%%%%%%%%%%%%%%
Therefore, the individual probability becomes:
%%%%%%%%%%%%%%%%%%%%%%%
\begin{eqnarray}
P_{\rm indv} &=& \dfrac{3v_i n_{\rm star} \Bigl(b_{\rm max}^2-b_{\rm min}^2 \Bigr)}{4 R_{\rm c}^3}\ln(R_{\rm c}/R_{\rm min}),\nonumber\\
&=& n_{\rm s}v_i\pi\Bigl(b_{\rm max}^2-b_{\rm min}^2 \Bigr)\ln(R_{\rm c}/R_{\rm min}),
\label{eq:Prb_ind}
\end{eqnarray}
%%%%%%%%%%%%%%%%%%%%%%%  
where, $R_{\rm min}$ is the lower radial cut-off. The analytical calculations treated the system of stars like a continuous `fluid', while the compact objects are really discrete, necessitating a $R_{\rm min}$ cut-off. We impose it in the following manner: considering $n_{\rm co}$ number of compact objects are distributed within radius $R_{\rm co}$ (which may be less than the radius of the whole cluster $R_{\rm c}$ for a dense core), $R_{\rm min}$ can be obtained from the following expression:
%%%%%%%%%%%%%%%%%%%%
\begin{equation}
\dfrac{4 \pi}{3}R_{\rm min}^3 n_{\rm co} \sim \dfrac{4 \pi}{3}  R_{\rm co}^3 \,,
\end{equation}
%%%%%%%%%%%%%%%%%%%%
which naively summarize the fact that total volume of the cluster is composed of  $n_{\rm co}$ number of spheres with radius $R_{\rm min}$. The cluster core, which matters the most here, is part of a larger system of stars, so it is reasonable to estimate the probability assuming the target source to be at the centre. Nevertheless, we analytically verified that if the target source is not at the centre of the sphere, the reduction in probability in \ref{eq:Prb_ind}, even very close to the edge of the sphere, is less than a few percent. Thus, considering the entire cluster, the probability of detectable hyperbolic interaction events becomes $P_{\rm clus}=n_{\rm star}P_{\rm indv}$, where the small scale structures of the cluster are ignored. 

We can now carry out a comparison between our framework and the usual scattering problem where the initial distance is assumed to be infinite. The latter is already employed to obtain the event rates for hyperbolic encounters inside a \gc~(\citealt{Capozziello:2008mn}) and between primordial BHs (\citealt{Garcia-Bellido:2017knh}). In the standard framework of hyperbolic scattering (e.g. gas molecules inside a closed container), where the perturber approaches from infinity, the rate of interaction per target is given by
%%%%%%%
\begin{equation}
 P_{\rm gm}=n\sigma v_{\rm m},
 \label{eq:Gas_Molecule}
\end{equation}
%%%%%%
where, $n$, $\sigma$ and $v_{\rm m}$ denote the number density of perturber, the cross-section for the interaction of interest, and the relative velocity between the perturber and the target respectively. Interestingly, the number densities and velocities can be easily mapped, while the logarithm term in \ref{eq:Prb_ind} has no counterpart in \ref{eq:Gas_Molecule}. Besides, the area of cross-section, $\sigma$ also resembles with the term $\pi(b_{\rm max}^2-b_{\rm min}^2 )$, where we note that the impact parameter $(b)$ in our model is mimicked by $r\sin\theta$, and area of cross section becomes $\sigma \approx \pi b^2$. In summary, we retrieve most of the components of the usual scattering classical problem, except additional logarithm terms containing information regarding the cluster's properties. Finally, we note that recently \citealt{Kocsis:2006hq} calculated the rate without taking into account the local effects. Albeit their setup is quite different from ours in the details, however, it is possible to compare our results at an order of magnitude level to highlight how local effects alter the cross section calculation. For example, with the typical parameters in the local Universe, we found that our rate without the log factor becomes $\mathcal{O}(10^{-2})$ for advanced LIGO which matches with their estimation. While it will be interesting to take into account more details of astrophysical distributions using our set up in future work, in this paper we limit ourselves to conservative effective values for astrophysical quantities (such as masses of the binary components and core radius) so that our results are scalable and easy to interpret.

Before taking into account different galaxies, we first estimate the detectable BH encounters rates for the Milky Way galaxy in the GCs (\citealt{Weatherford_2020}) and its bulge (\citealt{Olejak:2019pln}) and for the Andromaeda galaxy assuming similar numbers. The rates turned out to be very small, less than $10^{-7}$ per year. To incorporate the effects from galaxies at different redshifts, we consider the number density of the Milky Way Equivalent Galaxy (MWEG) to be $n(z)$, which can be assumed to be a constant here (\citealt{Abadie:2010cf}). At a comoving distance $r$, the number of galaxies between redshift $z$ to $z+dz$ is given as (\citealt{Ain:2015mea,Mazumder:2014fja,Baibhav:2019gxm})
%%%%%%%%%%%%%%%%
\begin{equation}
\mathcal{N}(z)=\dfrac{dN}{dz}=\dfrac{4\pi r^2}{H_0}\dfrac{c n(z)}{E(z)},
\label{eq:number}
\end{equation}
%%%%%%%%%%%%%%%%
where, $H_0$ is the Hubble constant, and Hubble parameter $E(z) = \sqrt{\Omega_\Lambda + (1+z)^3 \Omega_{\rm m}}$.
We use the base $\Lambda$CDM model for the calculations using standard cosmological parameters (\citealt{Aghanim:2018eyx}), $H_0 = 67.4$~km/s/Mpc, $\Omega_{\rm m}=0.315$ and $\Omega_\Lambda = 0.685$.
Finally, we arrive at the following expression to obtain the total event rate, accounting for cosmological time dilation with the the additional $1+z$ factor,
%%%%%%%%%%%%%%%%
\begin{equation}
P_{\rm tot}=\int^{z_{\rm max}}_{z_{\rm min}} (1+z)^{-1} P_{\rm clus} \mathcal{N}(z) n_{\rm gc}dz=\int^{z_{\rm max}}_{z_{\rm min}}\tau_{\rm tot}dz,
\label{eq:PTOTAL}
\end{equation}
%%%%%%%%%%%%%%%%
where, $n_{\rm gc}$ is the number of globular cluster in a galaxy.
We impose a redshift cut-off of $z_{\rm min}=0.0002$ (corresponding to $d_{\rm L} \sim 1$Mpc), contributions from redshits below that are  negligible to our results.

The final detectable event rates for different models and detectors are listed in \ref{Tab_01}. As these model deals with a number of parameters, we may need the distribution of its parameters to obtain the final rates. Since these distributions are uncertain and numerous possibilities are available in literature, bringing in those in the calculations will reduce the transparency of our estimate, without necessarily providing useful information. So we use average quantities as much as possible and use simplistic evolution profiles, staying on the conservative side. We first assume that the distribution of velocity in a cluster is isotropic, in which case the probability of a detectable encounter becomes proportional to $(\theta_{\rm max}^2 - \theta_{\rm min}^2)/4$. For simplicity, we assume that initially the velocities of all objects are of order the virial velocity (\ref{eq:virial_vel}) and that there is no anisotropy in the velocities. These assumptions should be of sufficient accuracy in the central regions of a star cluster; the central region is always the most dynamically relaxed and radial anisotropy grows as the distance from the center grows (e.g., \citealt{heggie_hut_2003}; \citealt{10.1093/mnras/stv2574}). Note that both are conservative estimates. Deviations from these assumptions, such as a preferred direction for the velocities, say towards the cluster centre, or the lower tail of the velocity distribution, would increase encounters. The latter is because, if the initial velocity of the compact objects is high, the event rate in time can go up (as individual probability $P_{\rm indv} \sim v_i$), but reduces the probability of collision as $\theta_{i}^2$ ($\sim 1/v_i^2$). Besides, there are trade-offs involved in the choice of mass as well. As for binary mergers, higher the mass, lower the frequency, but more the energy which increases the volume of detectability, especially for redshifted sources when observed with the next generation detectors.

\begin{figure}
    \centering
    \includegraphics[width=0.45\textwidth]{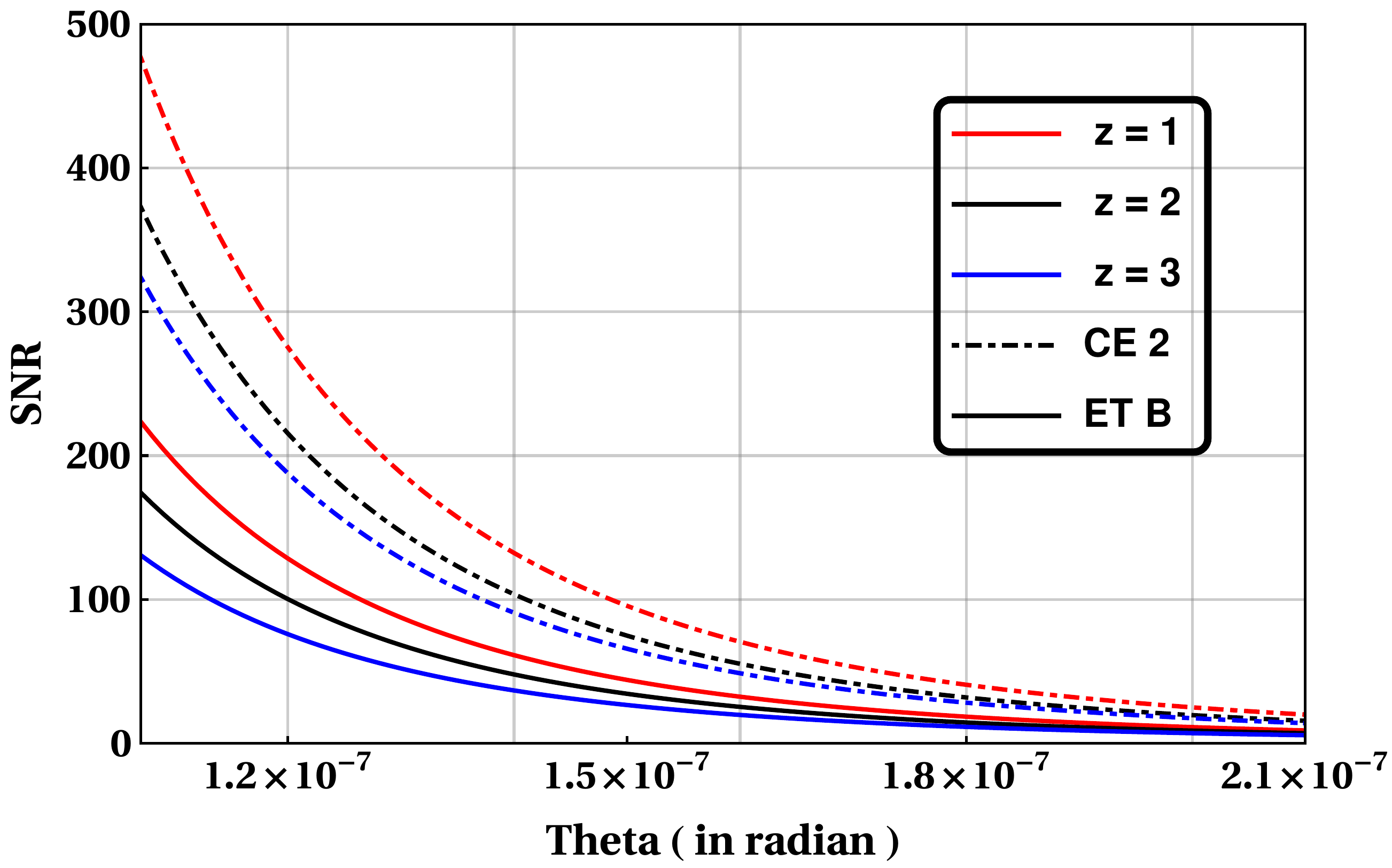}
    \includegraphics[width=0.42\textwidth]{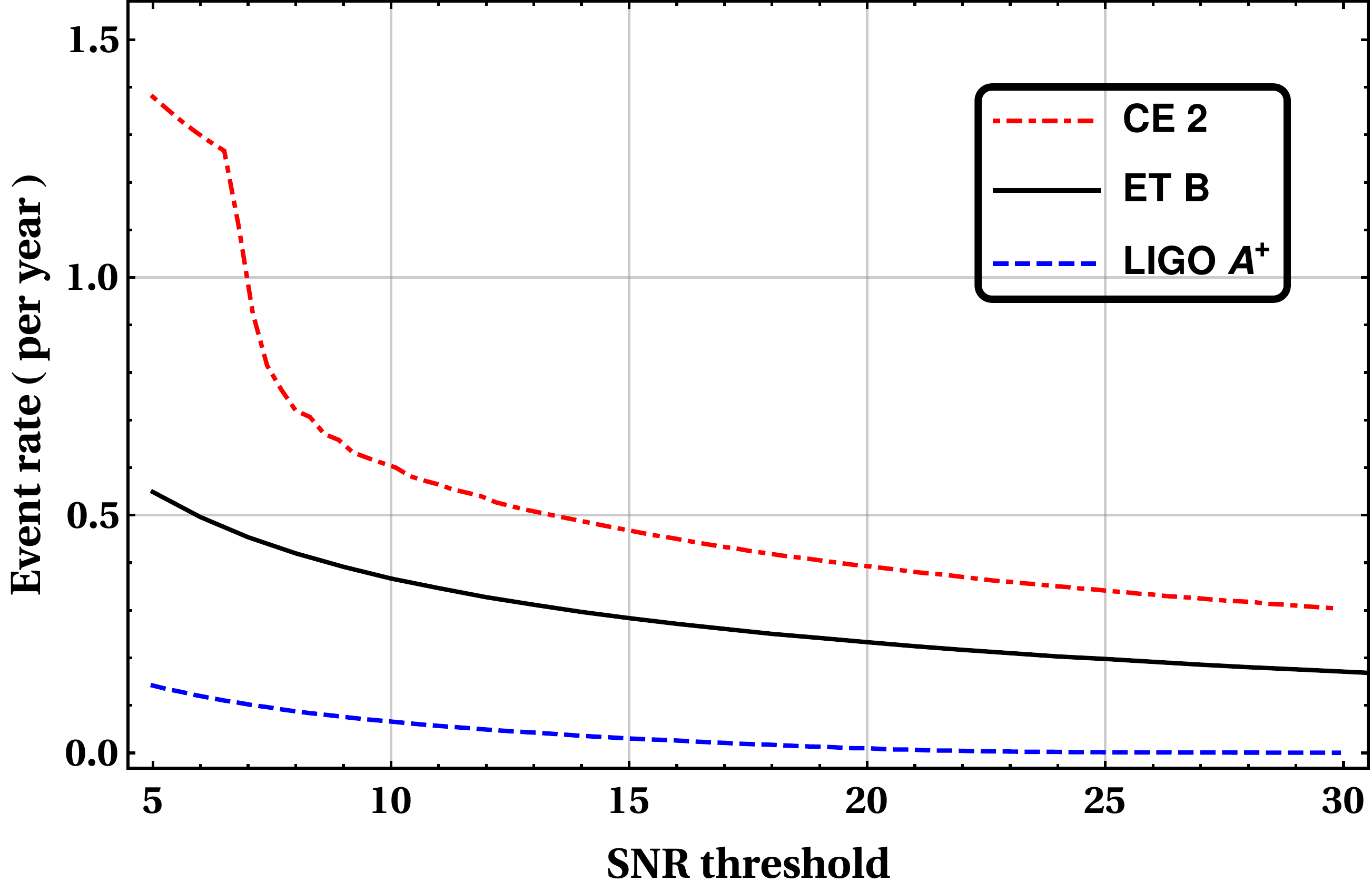}
    \caption{The upper panel shows the variation of SNR with respect to $\theta$ for different redshifts, while the lower panel depicts event rate as a function of SNR threshold. We set the initial parameters as, $r_i=0.5$pc, and $m_1=m_2=10 \Msun$. Detection probability is less for low SNR as well as small $\theta$, so the peak detection rate corresponds to a mid-value in the $\theta_{\rm min}$ to $\theta_{\rm max}$ range. As the top panel indicates, a large change in SNR is compensated by a minute change in $\theta_{\rm max}$ that determines the detection probability, leading to a weak dependence of the event rates on the SNR threshold. This also indicates why an order of magnitude more sensitive detector (with ten times more SNR) does not have thousand times more event rate. Interestingly, the kink in CE-2 event rate plot appears due to the sharp fall in source numbers at high redshift (implying an even sharper fall in comoving distance) in Model-III described below (left panel of \ref{fig:nbh_z}) and disappears if a smoother number density distribution in redshift is used.}
    \label{fig:SNR_Theta_Rate}
\end{figure}

Since the final detectable event rates come from the competition between a tiny probability of two-body encounter and large abundance of compact objects, especially near the core of the globular clusters, the parameters must be chosen carefully. Unfortunately though there is enormous uncertainly in literature about these numbers stemming mainly from the uncertainties in the distributions of initial cluster properties and formation times. In fact, both detection or non-detection of these events with GW observatories can significantly improve our understanding of these astrophysical parameter distributions. From an extensive set of literature, we gather that on the average, a GC has the following properties.
%%%%%%%%%%%%%%%%%%%%%%%%
\begin{itemize}

    \item They are typically of $\sim 10$pc radius, sometimes with an extended less dense halo, with up to a few million stars, (\citealt{gratton2019globular}) including white dwarfs, neutron stars and black holes. Their cores are dense, $\sim 10^5 / {\rm pc}^{3}$ stars (\citealt{Ivanova:2007bu,Weatherford_2020}), where a large number of interactions can take place.
    
    \item On an average, few $100$ BHs exist in GC cores up to a redshift of a few. A recent study found $\sim 50-100$ BHs in the core $\sim 0.5$pc region of the Milky Way GCs at the present epoch (\citealt{Weatherford_2020}) (where the authors acknowledge that they estimate a smaller number of BHs compared to other studies using different models), while the number was $\sim 1000$ at redshift $\sim 2$. There are also simulations that suggests $\sim 1000$ BHs of $\sim 20-100 \Msun$ may be retained in the GC over $12$~Gyr if the core is not very compact with a small fraction of the BHs forming binaries (\citealt{Morscher:2014doa}).
    
    \item Though $\sim 1$\% of the stellar population are NS (\citealt{camenzind2007compact}), could be more for GC which are relatively older (\citealt{Benacquista:2011kv}), only $\sim 30$\% of them may be retained due to their high natal kicks (\citealt{Kulkarni:1993fr}). Present simulations show that at present in the central  GC cores, there are $\sim 200-300$ NSs (\citealt{Ivanova:2007bu}), which seems to be not evolving significantly with time (\citealt{Kremer:2019iul}).

\end{itemize}
%%%%%%%%%%%%%%%%%%%%
%
The models we choose are well within these limits. For all the cases considered here, unless otherwise stated, we assume that each GC has a median $n_{\rm star} = 8 \times 10^5$ stars with average mass $\sim 1 \Msun$, that corresponds to a virial velocity of $10.69$~km/s. Out of these $n_{\rm star}$ stars, we consider hyperbolic interactions of only $n_{\rm co}$ compact objects. We estimate the final rates for the following distributions in the central core of $0.5$pc radius (\citealt{Kremer:2019iul}). Results are listed in respective columns of \ref{Tab_01} for single detector $\text{SNR} \ge 5$. Based on the official GWTC-2 catalog by the LIGO-Virgo-KAGRA collaboration\footnote{https://www.gw-openscience.org/eventapi/html/GWTC/}, and OGC-2 (\citealt{Nitz:2019hdf}), and assuming that templates for hyperbolic interactions will be created if there is a chance of detection (work in this direction has already started \citealt{Nagar:2020xsk,Nagar:2021gss,Albanesi:2021rby}), we believe that the threshold SNR of 5, which translates to network SNR of 7 for two detectors and more than 8 for three detectors, seems to be a reasonable choice. Besides, with sophisticated analysis techniques, the lowest detectable SNR is always reducing. Nevertheless, detection rate depends weakly on the choice of SNR cut-off, much weaker than $1/\text{SNR}^3$ which one might have intuitively thought. For example, for a single ET-B detector, the detection rates using cut-off SNR $5$ and $8$ are $0.55$ and $0.42$, respectively. This is because SNR drops highly nonlinearly with $\theta$, a significant change in SNR cut-off is compensated by a small change in $\theta_{\rm max}$ which determines the detection probability. This feature is demonstrated in \ref{fig:SNR_Theta_Rate} for different detectors, and different redshifts. For example, the SNR drops from $\sim 300$ to $\sim 10$, while the theta only changes from $1\times 10^{-7}$ to $2 \times 10^{-7}$.
%%%%%%%%%%%%%%%%%%%%
%%%%%%%%%%%%%%%%%%%%
This is why it is difficult to obtain a simple scaling relation between minimum detectable SNR and the detection volume or event rates. It can be seen in \ref{Tab_01} that even if the sensitivity between detectors is enhanced by an order of magnitude (e.g., from AdvLIGO to Cosmic Explorer), event rates do not scale as the cube of sensitivity. This would imply that the conclusion of the present study remains identical in case we use a SNR cut-off $10$ instead of $5$. 

Based on the typical set of values used here, we can provide a back of the envelop estimate for the event rates useful to get insight to the analysis at hand. In our study, we found that $\theta_{\rm max} \sim 10^{-7}$, and, by using~\ref{eq:Prb_ind} and $P_{\rm clus}=n_{\rm co} P_{\rm indv}$, we approximately obtain that total probability of encounters in a GC is $\sim 10^{-12}$ per year. We now assume that the number of GC per MWEG is $\sim 200$ (\citealt{Sedda:2020wzl}), and integrate over all the Milky Way Equivalent Galaxies (MWEG) upto a redshift of $3.5$. Finally, we obtain a total of $\sim 10^{10}$ number of galaxies (\citealt{Mazumder:2014fja}), and $\sim 10^{12}$ number of GCs, with partial reduction due to cosmological time dilation, and the event rate turns out to be $\sim$ few per year.

\begin{figure}
\centering
\includegraphics[width=0.49\textwidth]{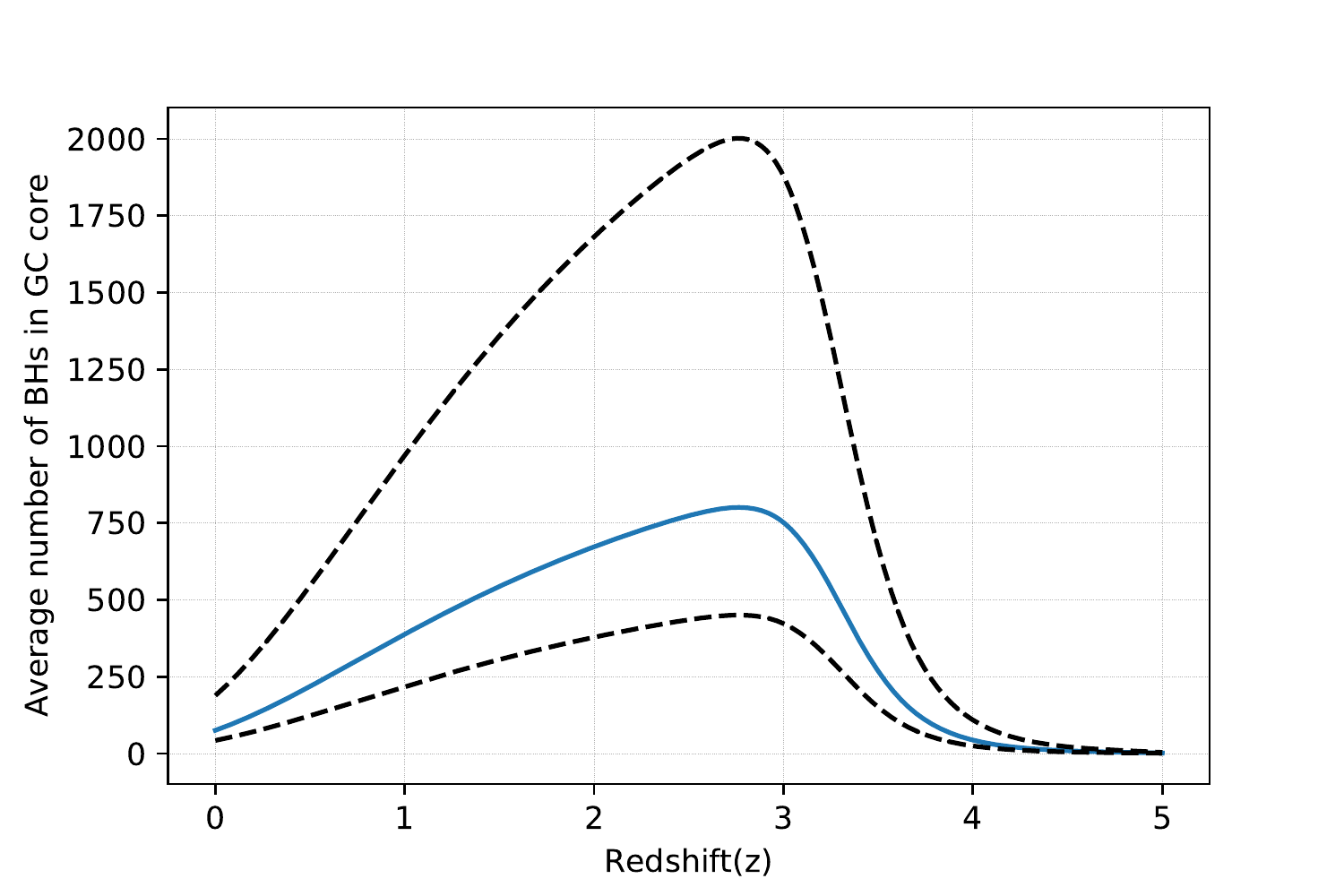}
\includegraphics[width=0.49\textwidth]{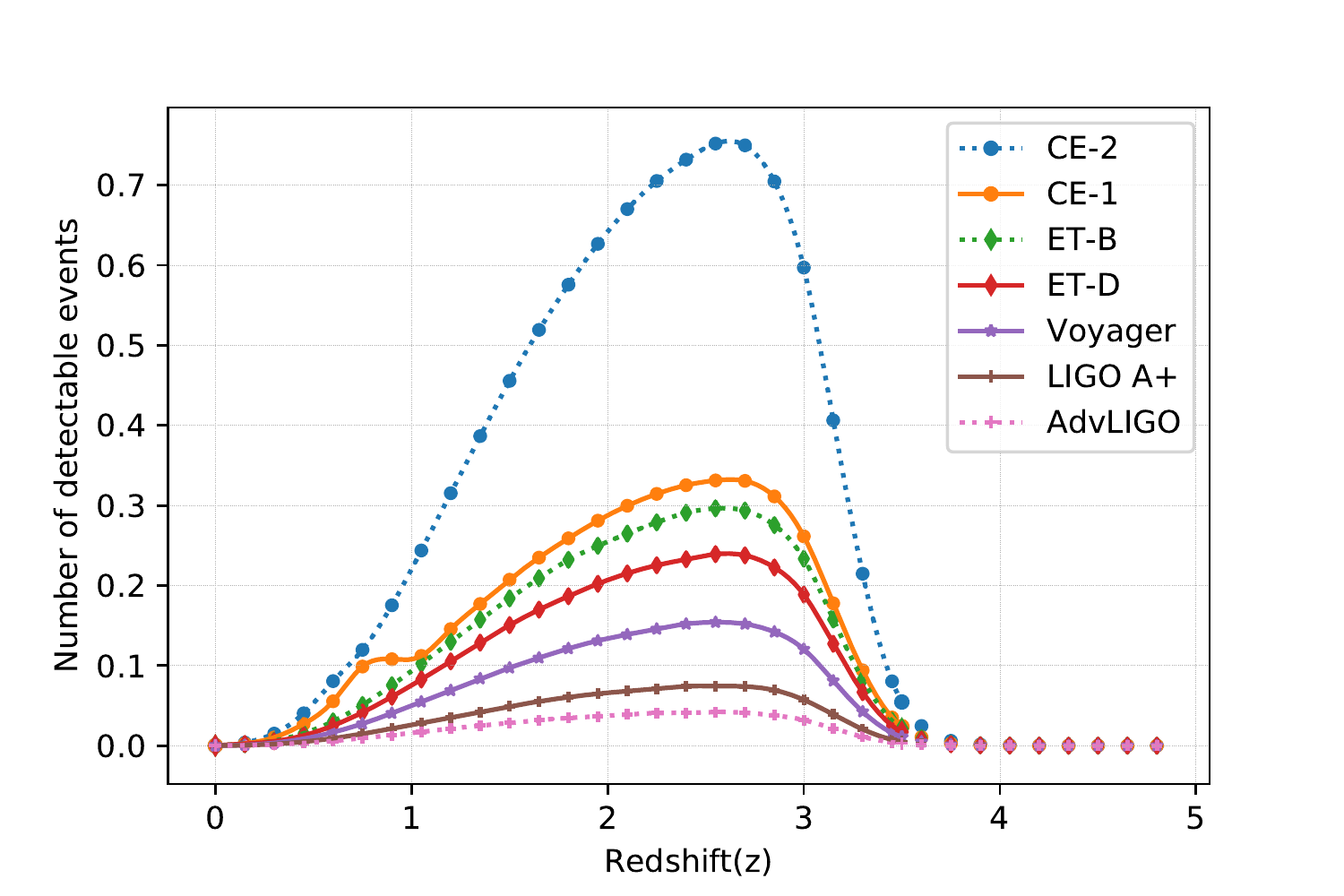}
\caption{In the upper panel, we have shown how the number of retained BHs varies over redshift inside the core of a \gc~for different models: the top, middle and bottom correspond to Model V, III, and IV respectively. At very high redshift, when the clusters are young, it is expected to follow cluster formation rate (with a subsequent delay in Supernova and blackhole formation). Roughly speaking, $10^{-3}N$ BHs form from $N$ stars and a fraction of these BHs are not readily ejected from the GC \citep{Breen:2013lca,Morscher:2014doa,Chatterjee:2016hxc}. At later times, BH ejection is driven by recoil from scattering which depends on the intricate interplay between scattering energies and cluster's global properties \citep{Rodriguez:2018rmd}. We roughly estimate a conservative time-varying number of BHs retained in the cluster based on these expectations. We assume, at $z=0$, the number of retained BHs is $\sim 80$ \citep{Weatherford_2020}, and the peak number is $\sim 800$ which is at $z \sim 3$, close to the peak of cosmological star formation rate as well as the formation times of typical Milky Way \gc s. We then use linear interpolation to obtain the number of BHs at a given redshift. In the lower panel: event rate is shown corresponds to Model-III for different detectors for the following parameters, $n(z)=0.0116~\text{Mpc}^{-3}$, $n_{\rm gc}=200$ \citep{Sedda:2020wzl}, $R_{\rm co}=0.5~\text{pc}$, and $m_1=m_2=10 \Msun$. In order to reproduce the above results, we assume that the initial distance to be $0.5$ pc. The small bump seen at lower redshifts in the curve for CE-1 was caused by a combined effect of frequency bandwidth and SNR threshold, and smoothens out if $f_{\rm min}$ is increased or SNR threshold is lowered.}
\label{fig:nbh_z}
\end{figure}

We now consider the following cluster models to estimate the expected event rates:
%%%%%%%%%%%%%%%%%%%%%
\begin{enumerate}

\item Model I: This is the simplest model, but useful, as the rates can be scaled as $n_{\rm co}^2 \log(n_{\rm co})$, other parameters and detector sensitivities remaining the same. Based on the discussions above (\citealt{Kocsis:2006hq,Morscher:2014doa}, we assume $1000$ BHs with an average mass of $10 \Msun$ which is fixed over $z$ and a redshift cutoff of $z_{\rm max}=3.5$ (lookback time of $\sim 12$Gyr).
    
    \item Model II: This model considers only NSs. From the studies mentioned above (\citealt{Kremer:2019iul,Ivanova:2007bu,Kulkarni:1993fr}), we take a total of $10$\% retention fraction for all formation channels, that is, $400$ NS in the core up to $z_{\rm max}=3.5$, which does not evolve significantly with lookback time. Note that, the retention rate may be even smaller. Furthermore, detecting NS encounters may get complicated due to significant tidal deformations. Hence, 
NS-NS encounters may be difficult to detect, unless future research indicates a higher abundance of NSs in GC cores.

    \item Model III: This we believe is the closest to reality. Although, as mentioned before, we adopt characteristic values of the important quantities while staying on the conservative side instead of full integration of profiles. Depending on the initial cluster property distribution, rate could be dominated by clusters residing in the high-density and young (the younger the cluster, the higher the black hole number) tail. Our formalism can easily accommodate such changes guided by detailed simulations in future. Here our goal is to motivate such studies. By considering the core to be composed of BHs only, based on a number of studies (\citealt{Kremer:2019iul,Rodriguez:2018rmd}), we assume that $0.1$\% of the stars became BH in a short period of time starting from a lookback time of $12$Gyr to $10$Gyr and $90$\% of them escaped (\citealt{Weatherford_2020}) from the GCs by the current epoch, as shown in Figure~\ref{fig:nbh_z}. We take an average mass of nominal $10 \Msun$, while adding NS population may increase the rates marginally.
    
    \item Model IV \& V: The rate estimation crucially hinges on the cluster's parameters. For accurate rate measurements, we need accurate distribution of these parameters. However, to obtain these detailed distributions, we need numerical modeling of a large number of star clusters in a large multi dimensional grid, which is challenging and also beyond the scope of this article. It is also hard to put constraints on this through observation. Instead we show how the estimates vary if we make different characteristic values of the parameters within the expected range exhibited by dense star clusters (e.g. \citealt{Kremer:2019iul}). In Model IV, we assume $v_i=7.561$~ Km/sec, and the variation of number of BHs with redshift is shown in \ref{fig:nbh_z} (the bottom curve). For Model V, we assume $v_i=15.122$~km/sec, and the retained BHs vs redshift is shown by the top curve in \ref{fig:nbh_z}. Here we have varied $v_i$ based on the value of $n_{\rm star}$ ($n_{\rm star} = 4\times 10^{5}$ for Model IV, and $n_{\rm star} = 1.6\times 10^{6}$ for Model V (\citealt{Kremer:2019iul})), as for the other parameter $R_{\rm co}$, the rates can be easily scaled as $1/R^3_{\rm co}$. From Model III to V, we have chosen $R_{\rm co}=0.5$~pc stimulated from the findings in \citealt{Kremer:2019iul}.
    
\end{enumerate}
%
%%%%%%%%%%%%%%%%%%%%%%%%
Note that in model II and III we considered NSs and BHs separately. These rates are additive and there will be NS-BH encounters as well, which are of the order few percent of BH-BH rates. Also, we did not account for primordial black holes (PBH) (\citealt{Garcia-Bellido:2017qal,Garcia-Bellido:2017knh}), which may significantly enhance the rates. It is worth mentioning that, since the number of interactions increase naively with the number of compact stars as $\sim n_{\rm co}^2 \ln(n_{\rm co})$, the denser clusters should dominate the rates, which may be underestimated by the average rates we are estimating here.
%%%%%%%%%%%%%%%%%%%%%%%%%%%%%%%
%%%%%%%%%%%%%%%%%%%%%%%%%%%%%%%%
%%%%%%%%%%%%%%%%%
%%%%%%%%%%%%%%%%%%%%%%%%%%%%%%%%
\begin{table*}
\centering
\begin{threeparttable}
\setlength{\tabcolsep}{1.5pt}
\caption[The.]{The following table contains the estimated event rates per year for different detectors (\citealt{ligodoc1,ligodoc2,CE:cite,Regimbau:2012ir}) for single detector SNR \tnote{\dag} $\ge 5$ (network $\rm{SNR} > 7$ for two detectors) in all the globular clusters in the Universe up to a redshift of $3.5$, $3.5$ and $5$ respectively. We assume an average comoving galaxy number density of $0.0116 \rm{Mpc}^{-3}$ MWEG (\citealt{Abadie:2010cf}). Rates are considerably affected by different mass values.  We have used mean virial velocity here (identical for each of the column, i.e., $v_i \cong 10.69$~km/sec), the lower tail of the velocity distribution may increase this number. The rates can increase significantly due to the density profile peaking at the centre. Note that the event rates are listed for a year, though {\em each of these facilities are expected to operate for at least $10$ years at different sensitivity levels and $4-5$ detectors could be operating at A+ like sensitivity in the next few years}, so the integrated detection rate is much higher. Though coincident detection in two observatories will be needed to claim a detection. In passing, we should note that event rates in different Galaxies like Milky Way or Andromeda, are extremely small, $\sim 10^{-8}$ to $10^{-7}$ per year, and less encouraging to pursue further.\\}
\label{Tab_01}
%\vskip 2mm
%%%%%%%%%%%%%%%%%%
\begin{tabular}{ccccccc}
 \hline\hline
 \multirow{5}{*}{Detector}& \multicolumn{5}{c}{Expected event rate (per year)} \\
     & Model I & Model II & Model III & Model IV & Model V\\
     & $n_{\rm co}=1000$  &
      $n_{\rm co}= 400$ & $n_{\rm co}$: \ref{fig:nbh_z} & $n_{\rm co}$: \ref{fig:nbh_z} & $n_{\rm co}$: \ref{fig:nbh_z}\\
     & $m=10 M_\odot$  & $m=2 M_\odot$ & $m=10 M_\odot$ & $m=10 M_\odot$ & $m=10 M_\odot$ \\
     & \citealt{Kocsis:2006hq} & \citealt{Kulkarni:1993fr} & \citealt{Weatherford_2020} & \citealt{Kremer:2019iul} & \citealt{Kremer:2019iul} \\
     & \citealt{Morscher:2014doa} & \citealt{Ivanova:2007bu} & \citealt{Kremer:2019iul} & & \\
     & & \citealt{Kremer:2019iul}& & &\\
   \hline
    Adv. LIGO  &  $0.29$ & $0.001$ & $0.08$ & $0.03$ & $0.39$\\
LIGO A+ &  $0.50$ & $0.002$ & $0.14$ & $0.06$ & $0.70$\\
Voyager   &  $0.98$ & $0.004$ & $0.20$ & $0.12$ & $1.44$\\
ET-D   & $1.51$ & $0.007$ & $0.44$  & $0.18$ & $2.23$\\
ET-B   & $1.85$ & $0.009$ & $0.55$  & $0.22$ & $2.78$\\
CE-1   & $2.34$ & $0.010$ & $0.63$ & $0.25$ & $4.14$ \\
CE-2   & $4.61$ & $0.014$ & $1.38$ & $0.35$ & $6.03$
\\ 
\hline\hline
\end{tabular}
%%%%%%%%%%%%%%%%%
\footnotesize
\begin{tablenotes}
\item[\dag] The expected rates scale rather weakly with any reasonable choice of SNR threshold. For instance, with single detector SNR threshold as high as $8$, the rate for CE-2 with Model~III changes from $1.38$/year to $1$/year.
\end{tablenotes}
\end{threeparttable}
%%%%%%%%%%%%%%%%%%
\end{table*}
%%%%%%%%%%%%%%%%%%
Finally, we should emphasize that the rates given in \ref{Tab_01} or shown in \ref{fig:nbh_z} correspond to the rates for detectable events, while the actual number of events can be much larger depending on $\theta_{\rm max}$ or $b_{\rm max}$. Even though hyperbolic interactions can have wide range of $b_{\rm max}$, many of them may not have any detectable effects. However, to estimate the event rate for total encounters, we need to specify $b_{\rm max}$. Here we are more concerned with close encounters which can have any chance of detection in the near future. Based on the cluster models, we estimate that $b_{\rm max}$ needs to be less than $\sim 10^{-1}$~AU to have any detectable impact. This is an optimistic choice, which corresponds to $r_{p} \sim 175 r_{\rm s}$. In practice, for events detectable with the present and next generation detectors, $r_{p}$ is few $r_{\rm s}$ only. However, such optimistic choice provide a crude estimate of how many hyperbolic interactions take place inside a cluster. With this, we obtain an event rate of $\sim 0.9$ $\text{GPC}^{-3}\text{yr}^{-1}$ independent of any detector.

Before closing this section, we should say a word about the uncertainties in the cluster's parameters, and how they affect the final rate estimation. In order to make a simplistic analysis, we assume that the number of retained BHs follows, $n_{\rm co}=n_{\rm o}f(z,\bar{M})$, where $n_{\rm o}$ is the maximum number of compact objects inside the cluster. The distribution function $f(z,\bar{M})$ depends on the redshift and average mass of the entire cluster $\bar{M}$. With this, we rewrite $P_{\rm clus}$ as,
%%%%%%%%%%
\begin{eqnarray}
    P_{\rm clus}&=&\dfrac{n^{2}_{\rm o}v_i\Big(b^2_{\rm max}-b^2_{\rm min}\Big)\Big[f(z,\bar{M})\Big]^2\ln(n_{\rm o}f(z))}{4 R^3_{\rm co}},\nonumber \\
    &=& \mathcal{G}_{\rm clus}g(z)\Big[f(z,\bar{M})\Big]^2\ln(n_{\rm o}f(z,\bar{M})),
\end{eqnarray}
%%%%%%%%%%
where $\mathcal{G}_{\rm clus}=(n^{2}_{\rm o}v_ib^2_{\rm min})/(4 R^3_{\rm co})$, is independent of the redshift parameter, and contains only cluster's parameters, and $g(z)=b^2_{\rm max}/b^2_{\rm min}-1$, depends on redshift, cluster, as well as the detector. In the expression $\mathcal{G}_{\rm clus}$, both $n_{\rm co}$ and $v_i$ depends on the bulk properties of the cluster. The other parameter $b_{\rm min}$ depends on $v_i$ as follows: 
%%%%%%%%%%%
\begin{equation}
    r_{p}=2 r_{\rm s}=L^2/\big(GM(1+e)\big)=b^2_{\rm min}v^2_i/\big(GM(1+e)\big),
\end{equation}
%%%%%%%%%%%
and as this will be the closest encounter, the orbit is likely to be parabolic or weakly hyperbolic, i.e., $e \approx 1$, and we have, $b_{\rm min} \cong 2\sqrt{GMr_{\rm s}}/v_i$. Finally, \ref{eq:PTOTAL} becomes
%%%%%%%%%%%%
\begin{align}
 P_{\rm total} &=\mathcal{G}_{\rm clus} \int^{z_{\rm max}}_{z_{\rm min}} (1+z)^{-1}g(z)[f(z,\bar{M})]^2\ln(n_{\rm o}f(z)) \mathcal{N}(z) n_{\rm gc}dz,\nonumber\\
 &=\mathcal{G}_{\rm clus}\ln(n_{\rm o}) \int^{z_{\rm max}}_{z_{\rm min}} (1+z)^{-1}g(z)[f(z,\bar{M})]^2 \mathcal{N}(z) n_{\rm gc}dz \nonumber\\
&+\mathcal{G}_{\rm clus}\int^{z_{\rm max}}_{z_{\rm min}} (1+z)^{-1}g(z)[f(z,\bar{M})]^2\ln(f(z,\bar{M})) \mathcal{N}(z) n_{\rm gc}dz. \nonumber 
 \\
\end{align}
%%%%%%%%%%%%
Due to the presence of $g(z)$ and $f(z,\bar{M})$, it is nearly impossible to decouple the redshift factor from the cluster's parameter. In fact the detectable rates crucially depends on $z$, and a simple scaling relations for the final rate in terms of the cluster's properties seems unlikely. However, if we consider the detector independent rates, we assume a fixed $b_{\rm max}$, and pull it outside the integration. To address $f(z,\bar{M})$, we need to specify $\bar{M}$ of the cluster. For typical MWEG with constant number density ($n(z)=0.0116 \rm{Mpc}^{-3}$), $n_{\rm gc}=200$, and cluster with $\bar{M}=8 \times 10^5 \Msun$, $v_i=5.9\times 10^{12}R^{-1/2}_{\rm c}$, $f(z,\bar{M})$ as given in \ref{fig:nbh_z}, we obtain 
%%%%%%%%%%
\begin{eqnarray}
P_{\rm total}\sim 6\times 10^{31}n^2_{\rm o}R^{-1/2}_{\rm c} b^2_{\rm max}\ln(n_{\rm o})/(4 R^3_{\rm co}).
\end{eqnarray}
%%%%%%%%%%
In order to obtain $P_{\rm total}$, we need to provide a $b_{\rm max}$, which serves as a cut-off. As already mentioned earlier, to have any detectable impact in the next few years, we need to have $b_{\rm max} \sim 10^{-1}$~AU. By using this cut off for Model III, we found a total event rate $\sim 10$ per year. Finally, we should again emphasize that the above is the total event rate, and not the detectable event rates which are given in \ref{Tab_01} for different detectors and different models. 
%%%%%%%%%%%%%%%%%%%%%%%%%%%%%%%%%%%%%%%%%%%%%%%%%%%%%%%%%%%%%
\section{Discussion}\label{sec:Discussion}
%%%%%%%%%%%%%%%%%%%%%%%%%%%%%%%%%%%%%%%%%%%%%%%%%%%%%%%%%%%%%
In the present article, we have discussed the detectability of hyperbolic encounters occurring inside \gc s by the ground based GW detectors. The prime objective of the study is two-folded; firstly, we have redefined the basic component of a scattering problem such that it includes the localized effects that appear inside a cluster due to its finiteness. This requires redefining the initial conditions of a hyperbolic encounter, which, in fact, changes various orbital entities. In the second part, we evaluate the event rates per year of these encounters as observed various detectors. Not only advanced LIGO and its ongoing upgrades, but also, we have considered future generation detectors to probe these events. While the underlying machinery remains fairly simple and non-relativistic, the obtained results are enough to convince the likelihood to detect these events, which may be possible even with few years of operations of the present ground based detectors with the ongoing upgrades~\footnote{\href{https://emfollow.docs.ligo.org/userguide/capabilities.html}{https://emfollow.docs.ligo.org/userguide/capabilities.html}} of the GW detectors (\citealt{Abbott:2020qfu}), LIGO, Virgo, KAGRA, and upcoming LIGO-India. It is remarkable how the competition between very low probability of impact and very large number of objects lead to a reasonable number of detectable events. However, we should be mindful that even if the detectable rates seem to be promising, the current search algorithms are not compatible to observe these encounters. To detect any glimpse of the scattering interactions in GW detectors, we require the accurate waveform model in relativistic regime. The present work may boost these searches.

It is worth noting that here we considered only close encounters. They reach velocities close to $c$ at a much larger periastron distance, compared to bound inspiraling orbits, hence can be many-fold brighter. Events with higher impact parameter (larger eccentricities) can become detectable if either the masses of the binary components or the initial velocity, exceeds the currently accepted limits on the parameters. Consideration of the full velocity distribution, instead of the average virial velocity we used here and the density profile of the cluster peaking at the centre, may significantly increase the number of detectable events. Moreover, it also seems appropriate to include the relativistic corrections in the event rates. As a result, both the momentum and velocity would increase. The increased momentum may boost the radiated energy which may results in a rise in the rates. In contrast to it, the velocity has an inverse effect on the rates as $\theta_i \sim 1/v_i$. Therefore, we may expect a competition between different entities. Altogether these effects may not be significant, though careful calculations are in order to make a reliable comment. We leave it as a future work.

We have used simplistic, but very realistic models here to conservatively estimate the detectable event rates. While extensive simulations, for different distributions of masses, velocities of compact stars, density profiles and evolution of a cluster, galaxy number density variation with redshift will provide rigorous estimates for each of those large combinations of models, as being done for binary mergers (\citealt{Ivanova:2007bu,Fragione:2018vty,Kremer:2019iul}), the present study provides a fresh and clean outlook to hyperbolic or parabolic encounter events, and at any circumstances, the study can be useful. Also, with considerable uncertainly in our present understanding of formation and evolution of stars and galaxies, it is not even clear if such a simulation can, in effect, be sufficiently conclusive. Effort should therefore be invested in developing strategies for detecting these exciting encounters, to distinguish them from resembling instrumental glitches (\citealt{Mukund:2016thr,BAHAADINI2018172}). Detection of these encounters, in conjunction with observations of binary mergers, can help us constrain characteristics of BH populations. Even in the case of no detection, such searches can help us to constrain the abundance of isolated compact stars in the universe, which may include primordial black holes, while higher detection rates can change our understanding of stellar population, which was the case for above $20\Msun$ black holes after GW detection started.
%%%%%%%%%%%%%%%%%%
%%%%%%%%%%%%%%%%%%%%%%%%%%%%%%%%
\section*{Acknowledgement}
%%%%%%%%%%%%%%%%%%%%%%%%%%%%%%%%
We greatly acknowledge Jan Steinhoff, Bhooshan Gadre and Georgios Lukes-Gerakopoulos for useful discussions and support. S. Mukherjee is thankful to Max Planck Institute for Gravitational Physics, Potsdam, for a warm hospitality during an academic visit there where a part of this work was carried out. S. Mitra and S. Mukherjee acknowledge support from the Department of Science and Technology (DST), India, provided under the Swarna Jayanti Fellowships scheme.  SC acknowledges support from the Department of Atomic Energy, Government of India, under project no. 12-R\&D-TFR-5.02-0200. Finally, all the authors are indebted to the computational resources of IUCAA. This paper has been assigned IUCAA preprint number IUCAA-02/2020 and LIGO document number LIGO-P2000389.
%%%%%%%%%%%%%%%%%%%%%%%%%%%%%%%%%%%%%%%%%%%%%%%%%%
\section*{Data Availability}
%%%%%%%%%%%%%%%%%%%%%%%%%%%%%%%%%%%%%%%%%%%%%%%%%%
In the present context, the authors have used publicly available data which are cited wherever possible.
%%%%%%%%%%%%%%%%%%%%%%%%%%%%%%%%%%%%%%
\bibliographystyle{mnras}
\bibliography{References.bib} % if your bibtex file is called example.bib

\begin{thebibliography}{}
\makeatletter
\relax
\def\mn@urlcharsother{\let\do\@makeother \do\$\do\&\do\#\do\^\do\_\do\%\do\~}
\def\mn@doi{\begingroup\mn@urlcharsother \@ifnextchar [ {\mn@doi@}
  {\mn@doi@[]}}
\def\mn@doi@[#1]#2{\def\@tempa{#1}\ifx\@tempa\@empty \href
  {http://dx.doi.org/#2} {doi:#2}\else \href {http://dx.doi.org/#2} {#1}\fi
  \endgroup}
\def\mn@eprint#1#2{\mn@eprint@#1:#2::\@nil}
\def\mn@eprint@arXiv#1{\href {http://arxiv.org/abs/#1} {{\tt arXiv:#1}}}
\def\mn@eprint@dblp#1{\href {http://dblp.uni-trier.de/rec/bibtex/#1.xml}
  {dblp:#1}}
\def\mn@eprint@#1:#2:#3:#4\@nil{\def\@tempa {#1}\def\@tempb {#2}\def\@tempc
  {#3}\ifx \@tempc \@empty \let \@tempc \@tempb \let \@tempb \@tempa \fi \ifx
  \@tempb \@empty \def\@tempb {arXiv}\fi \@ifundefined
  {mn@eprint@\@tempb}{\@tempb:\@tempc}{\expandafter \expandafter \csname
  mn@eprint@\@tempb\endcsname \expandafter{\@tempc}}}

\bibitem[\protect\citeauthoryear{Aasi et~al.}{Aasi
  et~al.}{2015}]{TheLIGOScientific:2014jea}
Aasi J.,  et~al., 2015, \mn@doi [Class. Quant. Grav.]
  {10.1088/0264-9381/32/7/074001}, 32, 074001

\bibitem[\protect\citeauthoryear{Abadie et~al.}{Abadie
  et~al.}{2010}]{Abadie:2010cf}
Abadie J.,  et~al., 2010, \mn@doi [Class. Quant. Grav.]
  {10.1088/0264-9381/27/17/173001}, 27, 173001

\bibitem[\protect\citeauthoryear{Abbott et~al.}{Abbott
  et~al.}{2016}]{Abbott:2016blz}
Abbott B.,  et~al., 2016, \mn@doi [Phys. Rev. Lett.]
  {10.1103/PhysRevLett.116.061102}, 116, 061102

\bibitem[\protect\citeauthoryear{Abbott et~al.}{Abbott
  et~al.}{2017a}]{Evans:2016mbw}
Abbott B.~P.,  et~al., 2017a, \mn@doi [Class. Quant. Grav.]
  {10.1088/1361-6382/aa51f4}, 34, 044001

\bibitem[\protect\citeauthoryear{Abbott et~al.}{Abbott
  et~al.}{2017b}]{TheLIGOScientific:2017qsa}
Abbott B.,  et~al., 2017b, \mn@doi [Phys. Rev. Lett.]
  {10.1103/PhysRevLett.119.161101}, 119, 161101

\bibitem[\protect\citeauthoryear{Abbott et~al.}{Abbott
  et~al.}{2017c}]{Abbott:2017gyy}
Abbott B.~P.,  et~al., 2017c, \mn@doi [Astrophys. J.]
  {10.3847/2041-8213/aa9f0c}, 851, L35

\bibitem[\protect\citeauthoryear{Abbott et~al.}{Abbott
  et~al.}{2019}]{LIGOScientific:2018mvr}
Abbott B.,  et~al., 2019, \mn@doi [Phys. Rev. X] {10.1103/PhysRevX.9.031040},
  9, 031040

\bibitem[\protect\citeauthoryear{Abbott et~al.}{Abbott
  et~al.}{2020a}]{Abbott:2020qfu}
Abbott B.,  et~al., 2020a, \mn@doi [Living Rev. Rel.]
  {10.1007/s41114-020-00026-9}, 23, 3

\bibitem[\protect\citeauthoryear{Abbott et~al.}{Abbott
  et~al.}{2020b}]{Abbott:2020uma}
Abbott B.,  et~al., 2020b, \mn@doi [Astrophys. J. Lett.]
  {10.3847/2041-8213/ab75f5}, 892, L3

\bibitem[\protect\citeauthoryear{Abbott et~al.}{Abbott
  et~al.}{2020c}]{Abbott:2020khf}
Abbott R.,  et~al., 2020c, \mn@doi [Astrophys. J. Lett.]
  {10.3847/2041-8213/ab960f}, 896, L44

\bibitem[\protect\citeauthoryear{Acernese et~al.}{Acernese
  et~al.}{2015}]{TheVirgo:2014hva}
Acernese F.,  et~al., 2015, \mn@doi [Class. Quant. Grav.]
  {10.1088/0264-9381/32/2/024001}, 32, 024001

\bibitem[\protect\citeauthoryear{Adhikari et~al.}{Adhikari
  et~al.}{2019}]{Adhikari:2019zpy}
Adhikari R.~X.,  et~al., 2019, \mn@doi [Class. Quant. Grav.]
  {10.1088/1361-6382/ab3cff}, 36, 245010

\bibitem[\protect\citeauthoryear{Aghanim et~al.}{Aghanim
  et~al.}{2020}]{Aghanim:2018eyx}
Aghanim N.,  et~al., 2020, \mn@doi [Astron. Astrophys.]
  {10.1051/0004-6361/201833910}, 641, A6

\bibitem[\protect\citeauthoryear{Ain, Kastha  \& Mitra}{Ain
  et~al.}{2015}]{Ain:2015mea}
Ain A.,  Kastha S.,   Mitra S.,  2015, \mn@doi [Phys. Rev. D]
  {10.1103/PhysRevD.91.124023}, 91, 124023

\bibitem[\protect\citeauthoryear{Albanesi, Nagar  \& Bernuzzi}{Albanesi
  et~al.}{2021}]{Albanesi:2021rby}
Albanesi S.,  Nagar A.,   Bernuzzi S.,  2021, \mn@doi [Phys. Rev. D]
  {10.1103/PhysRevD.104.024067}, 104, 024067

\bibitem[\protect\citeauthoryear{Aso, Michimura, Somiya, Ando, Miyakawa,
  Sekiguchi, Tatsumi  \& Yamamoto}{Aso et~al.}{2013}]{Aso:2013eba}
Aso Y.,  Michimura Y.,  Somiya K.,  Ando M.,  Miyakawa O.,  Sekiguchi T.,
  Tatsumi D.,   Yamamoto H.,  2013, \mn@doi [Phys. Rev. D]
  {10.1103/PhysRevD.88.043007}, 88, 043007

\bibitem[\protect\citeauthoryear{Bahaadini, Noroozi, Rohani, Coughlin, Zevin,
  Smith, Kalogera  \& Katsaggelos}{Bahaadini et~al.}{2018}]{BAHAADINI2018172}
Bahaadini S.,  Noroozi V.,  Rohani N.,  Coughlin S.,  Zevin M.,  Smith J.,
  Kalogera V.,   Katsaggelos A.,  2018, \mn@doi [Information Sciences]
  {https://doi.org/10.1016/j.ins.2018.02.068}, 444, 172

\bibitem[\protect\citeauthoryear{Baibhav, Berti, Gerosa, Mapelli, Giacobbo,
  Bouffanais  \& Di~Carlo}{Baibhav et~al.}{2019}]{Baibhav:2019gxm}
Baibhav V.,  Berti E.,  Gerosa D.,  Mapelli M.,  Giacobbo N.,  Bouffanais Y.,
  Di~Carlo U.~N.,  2019, \mn@doi [Phys. Rev. D] {10.1103/PhysRevD.100.064060},
  100, 064060

\bibitem[\protect\citeauthoryear{Banerjee, Baumgardt  \& Kroupa}{Banerjee
  et~al.}{2010}]{10.1111/j.1365-2966.2009.15880.x}
Banerjee S.,  Baumgardt H.,   Kroupa P.,  2010, \mn@doi [Monthly Notices of the
  Royal Astronomical Society] {10.1111/j.1365-2966.2009.15880.x}, 402, 371

\bibitem[\protect\citeauthoryear{Benacquista \& Downing}{Benacquista \&
  Downing}{2013}]{Benacquista:2011kv}
Benacquista M.~J.,  Downing J.~M.,  2013, \mn@doi [Living Rev. Rel.]
  {10.12942/lrr-2013-4}, 16, 4

\bibitem[\protect\citeauthoryear{Berry \& Gair}{Berry \&
  Gair}{2010}]{Berry:2010gt}
Berry C.~P.,  Gair J.~R.,  2010, \mn@doi [Phys. Rev. D]
  {10.1103/PhysRevD.82.107501}, 82, 107501

\bibitem[\protect\citeauthoryear{Breen \& Heggie}{Breen \&
  Heggie}{2013}]{Breen:2013lca}
Breen P.~G.,  Heggie D.~C.,  2013, \mn@doi [Mon. Not. Roy. Astron. Soc.]
  {10.1093/mnras/stt1599}, 436, 584

\bibitem[\protect\citeauthoryear{Abbott et~al.}{CE:}{}]{CE:cite}
\url{https://cosmicexplorer.org/ }

\bibitem[\protect\citeauthoryear{Camenzind}{Camenzind}{2007}]{camenzind2007compact}
Camenzind M.,  2007, Compact objects in astrophysics.
Springer

\bibitem[\protect\citeauthoryear{Capozziello \& De~Laurentis}{Capozziello \&
  De~Laurentis}{2008}]{Capozziello:2008mn}
Capozziello S.,  De~Laurentis M.,  2008, \mn@doi [Astropart. Phys.]
  {10.1016/j.astropartphys.2008.07.005}, 30, 105

\bibitem[\protect\citeauthoryear{Chatterjee, Rodriguez  \& Rasio}{Chatterjee
  et~al.}{2017}]{Chatterjee:2016hxc}
Chatterjee S.,  Rodriguez C.~L.,   Rasio F.~A.,  2017, \mn@doi [Astrophys. J.]
  {10.3847/1538-4357/834/1/68}, 834, 68

\bibitem[\protect\citeauthoryear{Cho, Gopakumar, Haney  \& Lee}{Cho
  et~al.}{2018}]{Cho:2018upo}
Cho G.,  Gopakumar A.,  Haney M.,   Lee H.~M.,  2018, \mn@doi [Phys. Rev. D]
  {10.1103/PhysRevD.98.024039}, 98, 024039

\bibitem[\protect\citeauthoryear{De~Vittori, Jetzer  \& Klein}{De~Vittori
  et~al.}{2012}]{DeVittori:2012da}
De~Vittori L.,  Jetzer P.,   Klein A.,  2012, \mn@doi [Phys. Rev. D]
  {10.1103/PhysRevD.86.044017}, 86, 044017

\bibitem[\protect\citeauthoryear{De~Vittori, Gopakumar, Gupta  \&
  Jetzer}{De~Vittori et~al.}{2014}]{DeVittori:2014psa}
De~Vittori L.,  Gopakumar A.,  Gupta A.,   Jetzer P.,  2014, \mn@doi [Phys.
  Rev. D] {10.1103/PhysRevD.90.124066}, 90, 124066

\bibitem[\protect\citeauthoryear{Dymnikova, Popov  \& Zentsova}{Dymnikova
  et~al.}{1982}]{dymnikova1982bursts}
Dymnikova I.,  Popov A.,   Zentsova A.,  1982, Astrophysics and Space Science,
  85, 231

\bibitem[\protect\citeauthoryear{Punturo et~al.}{ET:}{}]{ET:cite}
\url{http://www.et-gw.eu/ }

\bibitem[\protect\citeauthoryear{Flanagan \& Hughes}{Flanagan \&
  Hughes}{1998}]{Flanagan:1997sx}
Flanagan E.~E.,  Hughes S.~A.,  1998, \mn@doi [Phys. Rev. D]
  {10.1103/PhysRevD.57.4535}, 57, 4535

\bibitem[\protect\citeauthoryear{Fragione \& Kocsis}{Fragione \&
  Kocsis}{2018}]{Fragione:2018vty}
Fragione G.,  Kocsis B.,  2018, \mn@doi [Phys. Rev. Lett.]
  {10.1103/PhysRevLett.121.161103}, 121, 161103

\bibitem[\protect\citeauthoryear{Garcia-Bellido \& Nesseris}{Garcia-Bellido \&
  Nesseris}{2017}]{Garcia-Bellido:2017qal}
Garcia-Bellido J.,  Nesseris S.,  2017, \mn@doi [Phys. Dark Univ.]
  {10.1016/j.dark.2017.10.002}, 18, 123

\bibitem[\protect\citeauthoryear{García-Bellido \& Nesseris}{García-Bellido
  \& Nesseris}{2018}]{Garcia-Bellido:2017knh}
García-Bellido J.,  Nesseris S.,  2018, \mn@doi [Phys. Dark Univ.]
  {10.1016/j.dark.2018.06.001}, 21, 61

\bibitem[\protect\citeauthoryear{Gond\'an, Kocsis, Raffai  \& Frei}{Gond\'an
  et~al.}{2018}]{Gondan:2017wzd}
Gond\'an L.,  Kocsis B.,  Raffai P.,   Frei Z.,  2018, \mn@doi [Astrophys. J.]
  {10.3847/1538-4357/aabfee}, 860, 5

\bibitem[\protect\citeauthoryear{Gratton, Bragaglia, Carretta, D’Orazi,
  Lucatello  \& Sollima}{Gratton et~al.}{2019}]{gratton2019globular}
Gratton R.,  Bragaglia A.,  Carretta E.,  D’Orazi V.,  Lucatello S.,
  Sollima A.,  2019, The Astronomy and Astrophysics Review, 27, 8

\bibitem[\protect\citeauthoryear{Grobner, Jetzer, Haney, Tiwari  \&
  Ishibashi}{Grobner et~al.}{2020}]{Grobner:2020fnb}
Grobner M.,  Jetzer P.,  Haney M.,  Tiwari S.,   Ishibashi W.,  2020, \mn@doi
  [Class. Quant. Grav.] {10.1088/1361-6382/ab6be2}, 37, 067002

\bibitem[\protect\citeauthoryear{Heggie \& Hut}{Heggie \&
  Hut}{2003}]{heggie_hut_2003}
Heggie D.,  Hut P.,  2003, The Gravitational Million–Body Problem: A
  Multidisciplinary Approach to Star Cluster Dynamics.
Cambridge University Press, \mn@doi{10.1017/CBO9781139164535}

\bibitem[\protect\citeauthoryear{Heggie \& Rasio}{Heggie \&
  Rasio}{1996}]{Heggie:1995yk}
Heggie D.~C.,  Rasio F.~A.,  1996, \mn@doi [Mon. Not. Roy. Astron. Soc.]
  {10.1093/mnras/282.3.1064}, 282, 1064

\bibitem[\protect\citeauthoryear{Huerta et~al.}{Huerta
  et~al.}{2018}]{Huerta:2017kez}
Huerta E.,  et~al., 2018, \mn@doi [Phys. Rev. D] {10.1103/PhysRevD.97.024031},
  97, 024031

\bibitem[\protect\citeauthoryear{Ivanova, Heinke, Rasio, Belczynski  \&
  Fregeau}{Ivanova et~al.}{2008}]{Ivanova:2007bu}
Ivanova N.,  Heinke C.,  Rasio F.,  Belczynski K.,   Fregeau J.,  2008, \mn@doi
  [Mon. Not. Roy. Astron. Soc.] {10.1111/j.1365-2966.2008.13064.x}, 386, 553

\bibitem[\protect\citeauthoryear{Iyer et~al.}{Iyer
  et~al.}{2011}]{LIGOM1100296-v2}
Iyer B.,  et~al., 2011, Internal working note LIGO-M1100296-v2.
{Laser Interferometer Gravitational Wave Observatory (LIGO)}

\bibitem[\protect\citeauthoryear{Jakobsen, Mogull, Plefka  \&
  Steinhoff}{Jakobsen et~al.}{2021a}]{jakobsen2021gravitational}
Jakobsen G.~U.,  Mogull G.,  Plefka J.,   Steinhoff J.,  2021a, Gravitational
  Bremsstrahlung and Hidden Supersymmetry of Spinning Bodies (\mn@eprint
  {arXiv} {2106.10256})

\bibitem[\protect\citeauthoryear{Jakobsen, Mogull, Plefka  \&
  Steinhoff}{Jakobsen et~al.}{2021b}]{Jakobsen:2021smu}
Jakobsen G.~U.,  Mogull G.,  Plefka J.,   Steinhoff J.,  2021b, \mn@doi [Phys.
  Rev. Lett.] {10.1103/PhysRevLett.126.201103}, 126, 201103

\bibitem[\protect\citeauthoryear{Aso, Michimura, Somiya, Ando, Miyakawa,
  Sekiguchi, Tatsumi  \& Yamamoto}{KAG}{}]{KAGRA_cite}
\url{http://gwcenter.icrr.u-tokyo.ac.jp/en/ }

\bibitem[\protect\citeauthoryear{Kalogera et~al.,}{Kalogera
  et~al.}{2019}]{kalogera2019deeper}
Kalogera V.,  et~al., 2019, Deeper, Wider, Sharper: Next-Generation
  Ground-Based Gravitational-Wave Observations of Binary Black Holes
  (\mn@eprint {arXiv} {1903.09220})

\bibitem[\protect\citeauthoryear{Kocsis, Gaspar  \& Marka}{Kocsis
  et~al.}{2006}]{Kocsis:2006hq}
Kocsis B.,  Gaspar M.~E.,   Marka S.,  2006, \mn@doi [Astrophys. J.]
  {10.1086/505641}, 648, 411

\bibitem[\protect\citeauthoryear{Kremer et~al.,}{Kremer
  et~al.}{2020}]{Kremer:2019iul}
Kremer K.,  et~al., 2020, \mn@doi [Astrophys. J. Suppl.]
  {10.3847/1538-4365/ab7919}, 247, 48

\bibitem[\protect\citeauthoryear{Kulkarni, McMillan  \& Hut}{Kulkarni
  et~al.}{1993}]{Kulkarni:1993fr}
Kulkarni S.~F.,  McMillan S.,   Hut P.,  1993, \mn@doi [Nature]
  {10.1038/364421a0}, 364, 421

\bibitem[\protect\citeauthoryear{Iyer et~al.}{LIG}{}]{LIGO_India_cite}
\url{https://www.ligo-india.in/ }

\bibitem[\protect\citeauthoryear{Longair}{Longair}{2007}]{longair2007galaxy}
Longair M.~S.,  2007, Galaxy formation.
Springer Science \& Business Media

\bibitem[\protect\citeauthoryear{Mazumder, Mitra  \& Dhurandhar}{Mazumder
  et~al.}{2014}]{Mazumder:2014fja}
Mazumder N.,  Mitra S.,   Dhurandhar S.,  2014, \mn@doi [Phys. Rev. D]
  {10.1103/PhysRevD.89.084076}, 89, 084076

\bibitem[\protect\citeauthoryear{Misner, Thorne, Wheeler  et~al.}{Misner
  et~al.}{1973}]{misner1973gravitation}
Misner C.~W.,  Thorne K.~S.,  Wheeler J.~A.,   et~al., 1973, Gravitation.
Freeman, SanFrancisco

\bibitem[\protect\citeauthoryear{{Moody} \& {Sigurdsson}}{{Moody} \&
  {Sigurdsson}}{2009}]{2009ApJ...690.1370M}
{Moody} K.,  {Sigurdsson} S.,  2009, \mn@doi [\apj]
  {10.1088/0004-637X/690/2/1370}, \href
  {https://ui.adsabs.harvard.edu/abs/2009ApJ...690.1370M} {690, 1370}

\bibitem[\protect\citeauthoryear{Morscher, Pattabiraman, Rodriguez, Rasio  \&
  Umbreit}{Morscher et~al.}{2015}]{Morscher:2014doa}
Morscher M.,  Pattabiraman B.,  Rodriguez C.,  Rasio F.~A.,   Umbreit S.,
  2015, \mn@doi [Astrophys. J.] {10.1088/0004-637X/800/1/9}, 800, 9

\bibitem[\protect\citeauthoryear{Mukund, Abraham, Kandhasamy, Mitra  \&
  Philip}{Mukund et~al.}{2017}]{Mukund:2016thr}
Mukund N.,  Abraham S.,  Kandhasamy S.,  Mitra S.,   Philip N.~S.,  2017,
  \mn@doi [Phys. Rev. D] {10.1103/PhysRevD.95.104059}, 95, 104059

\bibitem[\protect\citeauthoryear{Nagar, Rettagno, Gamba  \& Bernuzzi}{Nagar
  et~al.}{2020}]{Nagar:2020xsk}
Nagar A.,  Rettagno P.,  Gamba R.,   Bernuzzi S.,  2020, \mn@doi [Phys. Rev. D]
  {10.1103/PhysRevD.103.064013}, 103

\bibitem[\protect\citeauthoryear{Nagar, Bonino  \& Rettegno}{Nagar
  et~al.}{2021}]{Nagar:2021gss}
Nagar A.,  Bonino A.,   Rettegno P.,  2021, \mn@doi [Phys. Rev. D]
  {10.1103/PhysRevD.103.104021}, 103, 104021

\bibitem[\protect\citeauthoryear{Nitz et~al.,}{Nitz
  et~al.}{2019}]{Nitz:2019hdf}
Nitz A.~H.,  et~al., 2019, \mn@doi [Astrophys. J.] {10.3847/1538-4357/ab733f},
  891, 123

\bibitem[\protect\citeauthoryear{O'Leary, Kocsis  \& Loeb}{O'Leary
  et~al.}{2009}]{OLeary:2008myb}
O'Leary R.~M.,  Kocsis B.,   Loeb A.,  2009, \mn@doi [Mon. Not. Roy. Astron.
  Soc.] {10.1111/j.1365-2966.2009.14653.x}, 395, 2127

\bibitem[\protect\citeauthoryear{Olejak, Belczynski, Bulik  \&
  Sobolewska}{Olejak et~al.}{2020}]{Olejak:2019pln}
Olejak A.,  Belczynski K.,  Bulik T.,   Sobolewska M.,  2020, \mn@doi [Astron.
  Astrophys.] {10.1051/0004-6361/201936557}, 638, A94

\bibitem[\protect\citeauthoryear{Punturo et~al.}{Punturo
  et~al.}{2010}]{Punturo:2010zza}
Punturo M.,  et~al., 2010, \mn@doi [Class. Quant. Grav.]
  {10.1088/0264-9381/27/8/084007}, 27, 084007

\bibitem[\protect\citeauthoryear{Regimbau et~al.}{Regimbau
  et~al.}{2012}]{Regimbau:2012ir}
Regimbau T.,  et~al., 2012, \mn@doi [Phys. Rev. D]
  {10.1103/PhysRevD.86.122001}, 86, 122001

\bibitem[\protect\citeauthoryear{Rodriguez \& Loeb}{Rodriguez \&
  Loeb}{2018}]{Rodriguez:2018rmd}
Rodriguez C.~L.,  Loeb A.,  2018, \mn@doi [Astrophys. J. Lett.]
  {10.3847/2041-8213/aae377}, 866, L5

\bibitem[\protect\citeauthoryear{Rodriguez, Chatterjee  \& Rasio}{Rodriguez
  et~al.}{2016}]{PhysRevD.93.084029}
Rodriguez C.~L.,  Chatterjee S.,   Rasio F.~A.,  2016, \mn@doi [Phys. Rev. D]
  {10.1103/PhysRevD.93.084029}, 93, 084029

\bibitem[\protect\citeauthoryear{Saketh, Vines, Steinhoff  \& Buonanno}{Saketh
  et~al.}{2021}]{saketh2021conservative}
Saketh M. V.~S.,  Vines J.,  Steinhoff J.,   Buonanno A.,  2021, Conservative
  and radiative dynamics in classical relativistic scattering and bound systems
  (\mn@eprint {arXiv} {2109.05994})

\bibitem[\protect\citeauthoryear{Sathyaprakash et~al.,}{Sathyaprakash
  et~al.}{2019}]{sathyaprakash2019multimessenger}
Sathyaprakash B.~S.,  et~al., 2019, Multimessenger Universe with Gravitational
  Waves from Binaries (\mn@eprint {arXiv} {1903.09277})

\bibitem[\protect\citeauthoryear{Sedda}{Sedda}{2020}]{Sedda:2020wzl}
Sedda M.~A.,  2020, \mn@doi [Commun. Phys.] {10.1038/s42005-020-0310-x}, 3, 43

\bibitem[\protect\citeauthoryear{Tiongco, Vesperini  \& Varri}{Tiongco
  et~al.}{2015}]{10.1093/mnras/stv2574}
Tiongco M.~A.,  Vesperini E.,   Varri A.~L.,  2015, \mn@doi [Monthly Notices of
  the Royal Astronomical Society] {10.1093/mnras/stv2574}, 455, 3693

\bibitem[\protect\citeauthoryear{Tiwari et~al.}{Tiwari
  et~al.}{2016}]{Tiwari:2015gal}
Tiwari V.,  et~al., 2016, \mn@doi [Phys. Rev. D] {10.1103/PhysRevD.93.043007},
  93, 043007

\bibitem[\protect\citeauthoryear{Trani, Spera, Leigh  \& Fujii}{Trani
  et~al.}{2019}]{Trani:2019nij}
Trani A.~A.,  Spera M.,  Leigh N. W.~C.,   Fujii M.~S.,  2019, ]
  {10.3847/1538-4357/ab480a}

\bibitem[\protect\citeauthoryear{Acernese et~al.}{VIR}{}]{VIRGO_cite}
\url{https://www.virgo-gw.eu/ }

\bibitem[\protect\citeauthoryear{Vines, Steinhoff  \& Buonanno}{Vines
  et~al.}{2019}]{Vines:2018gqi}
Vines J.,  Steinhoff J.,   Buonanno A.,  2019, \mn@doi [Phys. Rev. D]
  {10.1103/PhysRevD.99.064054}, 99, 064054

\bibitem[\protect\citeauthoryear{Weatherford, Chatterjee, Kremer  \&
  Rasio}{Weatherford et~al.}{2020}]{Weatherford_2020}
Weatherford N.~C.,  Chatterjee S.,  Kremer K.,   Rasio F.~A.,  2020, \mn@doi
  [The Astrophysical Journal] {10.3847/1538-4357/ab9f98}, 898, 162

\bibitem[\protect\citeauthoryear{Zevin, Samsing, Rodriguez, Haster  \&
  Ramirez-Ruiz}{Zevin et~al.}{2019}]{Zevin:2018kzq}
Zevin M.,  Samsing J.,  Rodriguez C.,  Haster C.-J.,   Ramirez-Ruiz E.,  2019,
  \mn@doi [Astrophys. J.] {10.3847/1538-4357/aaf6ec}, 871, 91

\bibitem[\protect\citeauthoryear{Aasi et~al.}{cci}{}]{ccite}
\url{https://www.ligo.caltech.edu/ }

\bibitem[\protect\citeauthoryear{Chatterjee, Rodriguez  \&
  Rasio}{lig}{a}]{ligodoc1}
\url{https://dcc.ligo.org/LIGO-T1800084/public }

\bibitem[\protect\citeauthoryear{lig}{lig}{b}]{ligodoc2}
\url{https://dcc.ligo.org/LIGO-T1800042-v4/public }

\makeatother
\end{thebibliography}
%%%%%%%%%%%%%%%%%%%%%%%%%%%%%%%%%%%%%%%%%%%%%%%%%%%
\appendix
%%%%%%%%%%%%%%%%%%%%%%%%%%%%%%%%%%%%%%%%%%%%%%%%%%%
%%%%%%%%%%%%%%%%%%%%%%%%%%%%%%%%%%%%%%%%%%%%%%%%%%%%
\section{Gravitational radiation}\label{sec:Radiation}
%%%%%%%%%%%%%%%%%%%%%%%%%%%%%%%%%%%%%%%%%%%%%%%%%%%%
Given the quadrupole moment of the binary components is given by $D^{ij}$, the power radiated due to gravitational wave is described as follows:
%%%%%%%%%%%%%%%%
\begin{equation}
P=-\dfrac{G}{45 c^5}<\dddot{D}^{ij}\dddot{D}_{ij}>,
\end{equation}
%%%%%%%%%%%%%%%%
where $G$ and $c$ are usual constants,
%and the symbol \enquote*{$<>$} represents a scalar product.
%Besides,
and the quadrupole moment $D_{ij}$ is defined as, $D_{ij}=\mu(3 x_i x_j-\delta_{ij}r^2)$ (\citealt{misner1973gravitation}). In \ref{fig:collage}, we demonstrate the radiation in time domain for a typical initial conditions such that $t=0$ coincides with the minimum distance, i.e., periapsis --- where the power becomes maximum and gives rise to a burst-like structure. With the initial angle becomes smaller, the orbit becomes nearly parabolic, and the interaction takes place in a shorter period of time (shown in \ref{fig:collage}). In \ref{fig:collage}, we demonstrate the gravitational waveform in the time domain for different perturbation components given by (\citealt{misner1973gravitation})
%%%%%%%%%%%%%%%
\begin{equation}
    h^{ij}=-\dfrac{2G}{d c^4}\Ddot{D^{ij}}
\end{equation}
%%%%%%%%%%%%%%%
where $d$ is the distance of the source from Earth. Note that, $h_{+}(t)$ and $h_{\times}(t)$ are defined as the plus and cross polarization of the perturbation respectively. At large initial time ($t_0$) limit, i.e, $t_0 \gg 1$, we can show that $h_{+}(-t_0) \approx h_{+}(t_0)$, whereas, $h_{\times}(-t_0)$ and $h_{\times}(t_0)$ has a difference in order of magnitude. This feature is not new, and already discussed in literature (\citealt{DeVittori:2014psa,Garcia-Bellido:2017knh}), stating that the cross-polarization has nonzero memory effect. Therefore, the present results are in consonance with the existing literature.

\begin{figure}
\centering
\includegraphics[width=0.45\textwidth]{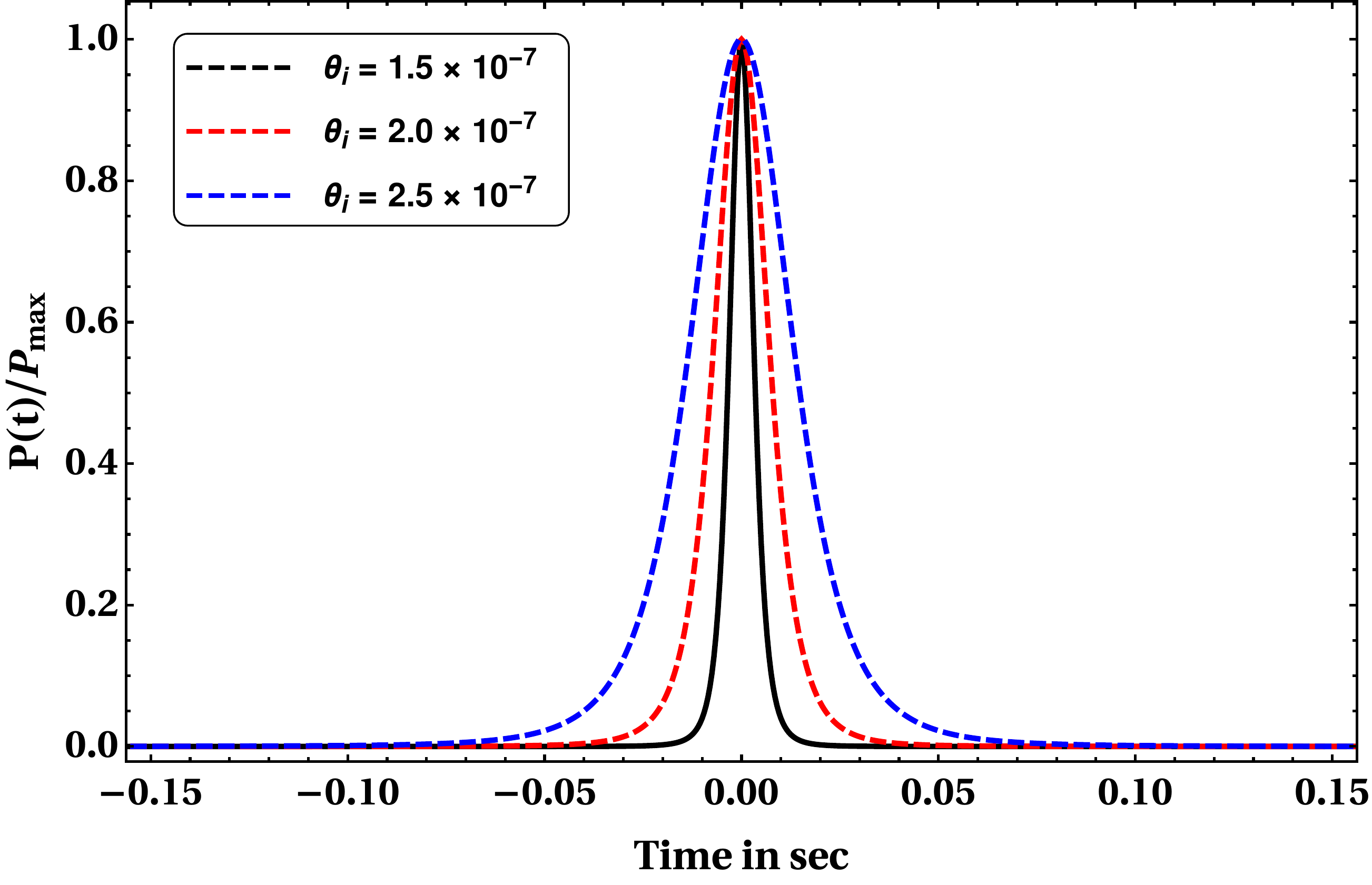}
\includegraphics[width=0.45\textwidth]{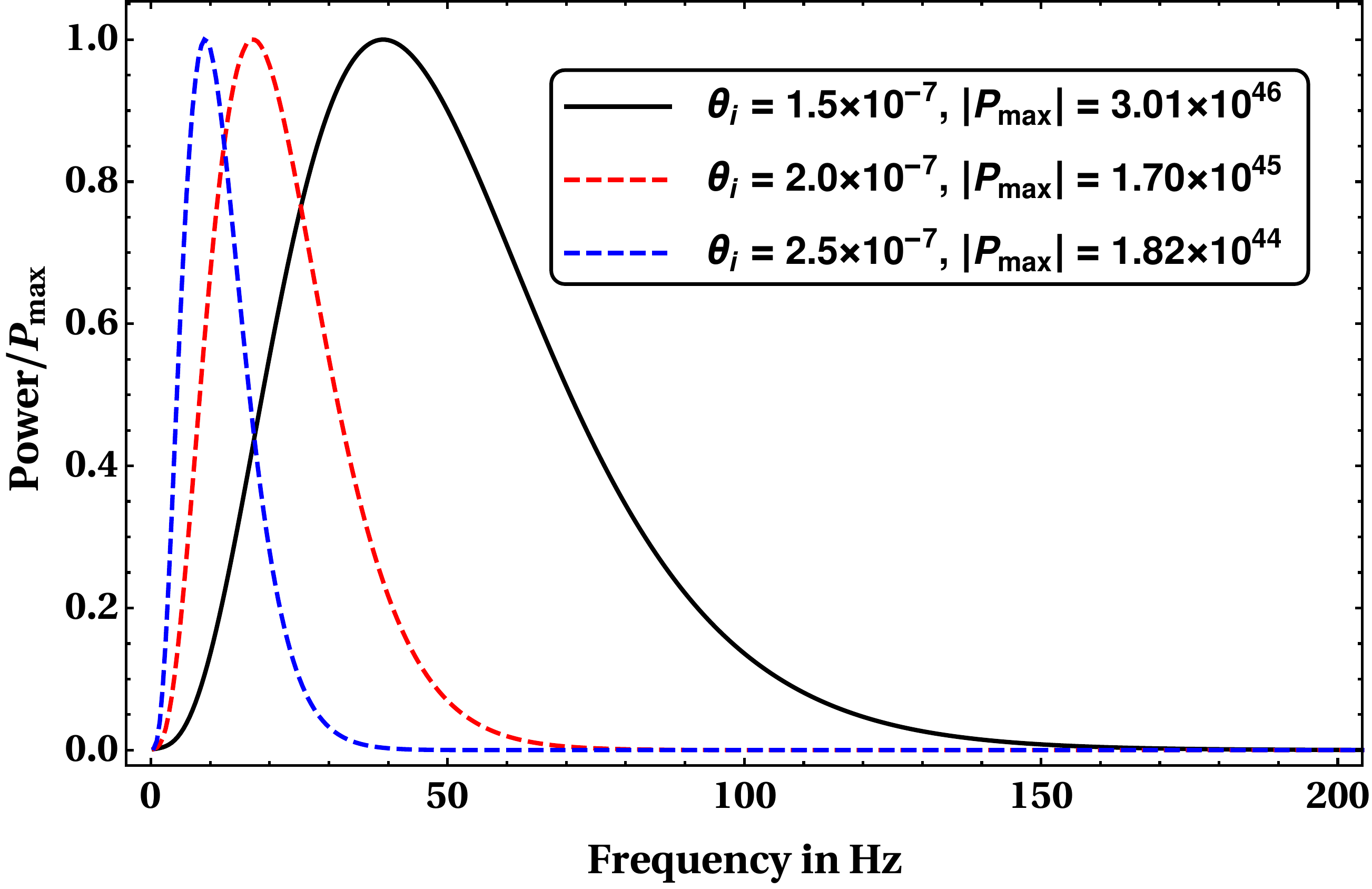}
\caption{In the above figure, we demonstrate the power radiation for different initial angle, while the initial distance is fixed at $r_i=1$pc, and $m_1=m_2=10 \Msun$. The plots are scaled with that maximum power $P_{\rm max}$ to appropriately highlight their differences. The corresponding distribution in the frequency domain is also shown.}
\label{fig:collage_appen}
\end{figure}

With the above calculations in time domain, we realize the primary components of the problem. However, in order to study the problem from detector's perspective, it is convenient to obtain the energy spectrum in frequency domain, which can be obtained via the Fourier transform. The following relation captures the relation between power radiation, $P(t)$, and energy spectrum, $dE/df$ in the frequency domain:
%%%%%%%%%%%%%%%%%%
\begin{equation}
\Delta E=\int^{\infty}_{-\infty} \dfrac{dE}{df}df=\int^{\infty}_{-\infty}\dfrac{dE}{dt}dt=\int^{\infty}_{-\infty}P(t)dt.
\end{equation}
%%%%%%%%%%%%%%%%%%
The above Fourier transformation was computed in literature for a different initial condition (\citealt{DeVittori:2012da,Garcia-Bellido:2017knh}). However, we serve our purpose with the numerical Fourier transform, and the spectrum is shown in \ref{fig:collage_appen}.
%%%%%%%%%%%%%%%%%%
%%%%%%%%%%%%%%%%%%
For a consistency check, we compare the above result with the parabolic limit as given in Ref. \citealt{Berry:2010gt} at large periapsis. We find an excellent accuracy within a numerical error $\sim$ $1 \%$. 
%%%%%%%%%%%%%%%%%%%%%%%%%%%%%%%%%%%%%%%%%%%%%%%%%%%%%%%%%%%%%%%%%%%%%%
%%%%%%%%%%%%%%%%%%%%%%%%%%%%%%%%%%%%%%%%%%%%%%%%%%%%%%%%%%%%%%%%%%%%%%
\section{Detectors \& signal to noise ratio}\label{sec:SNR}
%%%%%%%%%%%%%%%%%%%%%%%%%%%%%%%%%%%%%%%%%%%%%%%%%%%%%%%%%%%%%%%%%%%%%%
Our aim is to estimate the detection rates for 
for the current and upcoming detectors,
Advanced LIGO (\citealt{TheLIGOScientific:2014jea}),
Virgo (\citealt{TheVirgo:2014hva}),
KAGRA (\citealt{Aso:2013eba}),
LIGO-India (\citealt{LIGOM1100296-v2}),
LIGO-Voyager (\citealt{Adhikari:2019zpy}), 
Einstein Telescope (ET) (\citealt{Punturo:2010zza}) and
Cosmic Explorer (CE) (\citealt{Evans:2016mbw}).
%%%%%%%%%%%%%%%%
\begin{equation}
\rho^2=4\int^{f_{\rm  max}}_{{f_{\rm min}}}\dfrac{[\tilde{h}(f)]^2}{ S_{\rm h}(f)},
\end{equation}
%%%%%%%%%%%%%%%%
where, $\tilde{h}(f)$ is the Fourier transform of the gravitational wave signal computed at the detector, and $S_{\rm h}(f)$ is the one-sided noise spectral density, which are generally publicly available (\citealt{ligodoc1,ligodoc2,CE:cite,Regimbau:2012ir}). Given that the above expression is general and depends on various parameters involving the location of the source, it may be possible to further simply it for convenience. Following standard references (\citealt{Flanagan:1997sx}), We may consider the RMS average of the SNR, and finally have,
%%%%%%%%%%%%%%
\begin{eqnarray}
\rho_{\rm rms}^2 &=& \int^{f_{\rm  max}}_{{f_{\rm min}}} \dfrac{[h_{\rm c}(f(1+z))]^2}{5 f^2 S_{\rm h}(f)}df, \nonumber \\
 h_{\rm c}(f_{\rm e})&=&\dfrac{1+z}{\pi d_{\rm L}}\sqrt{\dfrac{2G}{c^3} \dfrac{dE_{\rm e}}{df_{\rm e}}},
\label{eq:rms_LIGO}
\end{eqnarray}
%%%%%%%%%%%%%%
where the factor of $1/5$ appears as an average of the detector's antenna functions in Advanced LIGO, $f_{\rm e}=(1+z)f$ denotes the frequency in the source frame, and $d_{\rm L}=(1+z)d$ is the \textit{Luminosity distance}. To furnish the same task for next generation detectors the expression for $\rho_{\rm rms}$ may subject to change. In the case of CE, the antenna functions are identical to AdvLIGO and $\rho_{\rm rms}$ is given by \ref{eq:rms_LIGO}; while, the noise profile is adequately better (\citealt{ligodoc1,CE:cite}). However, the same is not true for ET, where the detector's functions are given as follows (\citealt{Regimbau:2012ir}):
%%%%%%%%%%%%%%%%
%%%%%%%%%%%%%%%%%
\begin{eqnarray}
F_{+}&=&-\dfrac{\sqrt{3}}{4} \Bigl[(1+\cos^2\theta)\sin2\phi \cos2\psi \nonumber \\
 && \quad +2 \cos\theta \cos2\phi\sin2\psi\Bigr], \nonumber \\
F_{-}&=&\dfrac{\sqrt{3}}{4} \Bigl[(1+\cos^2\theta)\sin2\phi \sin2\psi, \nonumber \\
& &  \quad -2 \cos\theta \cos2\phi \cos2\psi\Bigr],
\end{eqnarray}
%%%%%%%%%%%%%%%%%
with $\theta,\phi$ denote the location on the sky, and $\psi$ is the polarization angle. In order to replicate \ref{eq:rms_LIGO} for ET, it is essential to find the average of above expressions over these angels (\citealt{Flanagan:1997sx}):
%%%%%%%%%%%%%%%%%%
\begin{equation}
F_{\pm \rm ave}^2=\dfrac{1}{4\pi}\int^{\theta=\pi}_{\theta=0} \int^{\phi=2\pi}_{\phi=0} d\Omega_{\theta,\phi}\int^{\psi=\pi}_{\psi=0} \dfrac{d\psi}{\pi}F_{\pm}^2=3/20.
\end{equation}
%%%%%%%%%%%%%%%%%%
Therefore, \ref{eq:rms_LIGO} now becomes
%%%%%%%%%%%%%%%%
\begin{equation}
\rho_{\rm rms}^2=\int^{f_{\rm  max}}_{{f_{\rm min}}} \dfrac{3[h_{\rm c}(f)]^2}{20 f^2 S_{\rm h}(f)},
\label{eq:rms_ET}
\end{equation}
%%%%%%%%%%%%%%%%
where, the expression for $h_{\rm c}(f)$ remains identical. Both \ref{eq:rms_LIGO} and \ref{eq:rms_ET} would be useful to constraint $\theta_i$, and obtain the event rate.
%%%%%%%%%%%%%%%%%%%%%%%%%%%%%%%%%%%%%%%%%%%%%%%

%%%%%%%%%%%%%%%%%%%%%%%%%%%%%%%%%%%%%%%%%%%%%%%%%%

%%%%%%%%%%%%%%%%% APPENDICES %%%%%%%%%%%%%%%%%%%%%

%%%%%%%%%%%%%%%%%%%%%%%%%%%%%%%%%%%%%%%%%%%%%%%%%%

%\bsp	% typesetting comment
\label{lastpage}
\end{document}